\documentclass[lineno]{jfm}

\usepackage{graphicx}
\usepackage{newtxtext}
\usepackage{newtxmath}
\usepackage{natbib}
\usepackage{hyperref}
\usepackage{tikz} 
\usepackage{subcaption}

\newcommand{\RomanNumeralCaps}[1]

\title{Shape considerations in wave-driven propulsion}

\author{Daire O'Donovan\aff{1}
  \corresp{\email{daire.odonovan@ucdconnect.ie}},
 \and Graham P. Benham\aff{2}}

\affiliation{\aff{1}School of Mathematics and Statistics, University College Dublin, Belfield, Dublin 4, Ireland
\aff{2}Department of Mechanical Engineering, University College London, Gower Street, London WC1E 6BT, UK.}

\begin{document}
\maketitle

\begin{abstract}
    At the fluid surface, an asymmetric wave field generated by an oscillating body produces thrust in what is known as wave-driven propulsion (WDP). Although this thrust is directed forward, the vertical oscillations inevitably produce waves radiating outwards in all directions. The effect of these side waves is not accounted for in simplified models focused on the fore--aft wave field. In this work, we take a two-dimensional shallow water approach, deriving thrust and power expressions that account for surface waves emitted in any direction. Supported by an analytically driven approach, we pose a numerical optimisation problem to maximise thrust, imposing appropriate Sommerfeld boundary conditions across subcritical and supercritical velocity regimes.
	The two-dimensional set-up allows us to understand how the side waves emitted by the body contribute to thrust and efficiency. Furthermore, we explore how maximal thrust depends on the shape of the body beginning with a square followed by ellipses of varying aspect ratios. Using these learnings, we include the shape of the body as a control variable to demonstrate how an optimised shape can lead to further thrust and efficiency gains.  
\end{abstract}

\section{Introduction}\label{sec1:intro}

Horizontal propulsion of a body at the fluid surface is a problem with a long and continued interest with a wide range of applications including biolocomotion \citep{Bush2006review}, water sports \citep{Dode2022} and ship hydrodynamics \citep{newman_marine_1977}. 
Recent works show a growing interest in leveraging the waves produced by a body to aid propulsion \citep{Benham2024propulsion,Rhee2022surferbot,Harris2025_review}. \citet{Benham2022gunwale} explored generating a thrust by jumping up and down on a canoe to induce an asymmetric wave field in what is known as gunwale bobbing. 
This example is the focus of the work to follow which we refer to as wave-driven propulsion (WDP). WDP is not limited to a canoe-sized body oscillating at the water surface. 
At smaller scales, WDP is used to describe bouncing and walking droplets \citep{Pucci2015floating_drops,Bush2006review} and capillary surfers or spinners \citep{Ho2023capillary_surfers,Barotta2023spinners,oza_theoretical_2023,oza_vertical_2026} achieving propulsion on a vibrating bath. 
At a similar scale, a honeybee confined to the fluid surface utilises WDP by beating its wings to generate an asymmetric wave field \citep{Roh2019bee}, inspiring the design of SurferBot \citep{Rhee2022surferbot}.
In another form of WDP, a young water strider was found to generate insufficient thrust from the surface waves produced (Denny's paradox). The observation of subsurface vortices that add further thrust resolved this apparent contradiction \citep{Hu2003water_strider,Bush2006review}. 
Furthermore, a three-dimensional approach was required to reconcile the varying ratios of wave and vortex propulsive strength found from differing modelling perspectives \citep{GAO_FENG_2011,BUHLER_2007,Steinmann2021}.

Focusing on WDP, the primary mechanism at play is to exploit the excess flow of momentum in the presence of waves as defined by \citet{LONGUETHIGGINS1964}, where the former went on to demonstrate the propulsive abilities of a wave-inducing raft \citep{Longuet-Higgins1977meanforces}. The vertical flapping or bobbing generates a wave field and the excess flow of momentum carried by the waves is proportional to the wave amplitude squared. To leading order in the shallow limit, the net thrust due to WDP is \citep{Longuet-Higgins1977meanforces}
\begin{equation}\label{eq:LH_thrust}
	F_{T} = -\frac{1}{2}\rho g\Big[h^2\Big]^{x^+}_{x^-}, 
\end{equation}
where $h$ is the wave amplitude, $\rho$ is density, $g$ is acceleration due to gravity, and $x^{\pm}$ represent the upstream and downstream wave field, respectively. Thus, a fore--aft asymmetry produces net forward momentum. 

In a one-dimensional shallow water set-up, our previous work \citep{odonovan2026} focuses on characterising the optimal contribution to WDP from the surface gravity waves generated by a periodically oscillating body. The time-averaged thrust is derived directly from the linearised shallow water equation to arrive at an equivalence relation between the thrust injected by the source and the thrust radiated away by the waves similar to \eqref{eq:LH_thrust}. The optimal control problem is divided into two separate cases where either the norm of the source term or the power injected is bounded for regularisation. To maximise thrust and efficiency, no wave is emitted upstream by the body. This is achieved for all velocities in the bounded power case, but only at harmonics dependent on dimensionless velocity and length in the bounded norm case.
This work is limited by the one-dimensional set-up, corresponding to a source producing waves purely in the direction of travel ($x$). 
However, all the WDP examples provided also produce waves travelling with a perpendicular vector component ($y$) which is not modelled. Therefore, to study a more realistic WDP set-up, we must include both $x$ and $y$ dynamics in our model to ultimately get closer to reality.

Modelling a two-dimensional surface will lead to multiple modelling questions to be addressed in the context of WDP. 
For example, a body travelling along the fluid surface is expected to produce a wedge shaped wake pattern \citep{rabaud2013_kelvin_or_mach,keeler2025_wake,Yuan2021ducks}. However, the wake of an oscillating body is not restricted to such a wedge when translating slower than the waves it produces (subcritical) (e.g. SurferBot). Instead, the body emits waves in all directions, but this wave field undergoes a Doppler shift in the moving frame of the body. Finding boundary conditions that radiate the wave out of the truncated numerical domain is not as simple as the one-dimensional set-up \citep{odonovan2026}. We are required to calculate an effective wavenumber for the outgoing wave at every position at the boundary for a given angle and velocity.  
However, when moving quicker than the emitted waves (supercritical), the oscillating body is restricted to a wedge-shaped wake due to causality. This is a shallow water, velocity-dependent wake, rather than a fixed angle Kelvin wake \citep{rabaud2013_kelvin_or_mach,keeler2025_wake}. 

These wave fields can only be observed by including both $x$- and $y$-directions, from which we are able to understand the contribution of side waves emitted from the body in WDP. Efficiency is expected to suffer since the component of the wave flux in the $y$-direction does not contribute to thrust in the $x$-direction.
We can then understand the role of body shape in WDP and controlling these side waves. The magnitude that this affects optimisation is then explored in the context of our body using WDP. We can draw inspiration from studies focused on drag \citep{boucher2018_thin,Benham2019_axisymmetric_bodies}, but focus purely on maximising thrust due to WDP for circular and elliptical bodies (representing boats) as well as a square body resembling SurferBot \citep{Rhee2022surferbot}. The thrust and efficiency outputs will be intrinsically linked to the control of side waves, helping us to understand preferable shapes and aspect ratios in optimisation. 

In the work to follow, we will use a linearised two-dimensional shallow water set-up, a minimal model that incorporates the effects mentioned while remaining simple enough to interpret and make analytical progress. In this model where surface tension, viscosity and vorticity are neglected, we focus purely on the surface wave field and its role in WDP. It permits us to define the thrust in terms of the wave amplitude squared (as in \eqref{eq:LH_thrust}) with terms contributing to both the fore--aft and side waves in this more realistic set-up. We can then evaluate the maximal thrust for elliptical sources of varying aspect ratios showing that longer bodies link to thrust while wider bodies link to efficiency.

We will begin with the two-dimensional shallow water equation and derive the power and thrust due to WDP in Section \ref{sec2:maths_setup}. The optimal control problem will begin with the start-up case, where velocity is zero, using both numerical and analytical approaches to validate the method for a square source in Section \ref{sec3:startup}. In Section \ref{sec4:subcrit}, we model the subcritical case with a circular body moving with velocity less than the wave speed, generalising the Sommerfeld boundary conditions. 
Following this, we consider an elliptical source, studying the impact of varying the aspect ratio on maximal thrust.
In Section \ref{sec6:supercrit}, we carry out the same investigation for velocities greater than the wave speed, the supercritical case, after which we demonstrate the potential for further improvements by including the shape as an additional control to optimise in Section \ref{sec:optimal_shaping}. To conclude we discuss the limitations and potential future directions in Section \ref{sec:discuss}.

\section{The Mathematical Set-up}\label{sec2:maths_setup}

\begin{figure}
	\centering
  \begin{tikzpicture}
		\node at (0,0) {\includegraphics[width=0.75\textwidth]{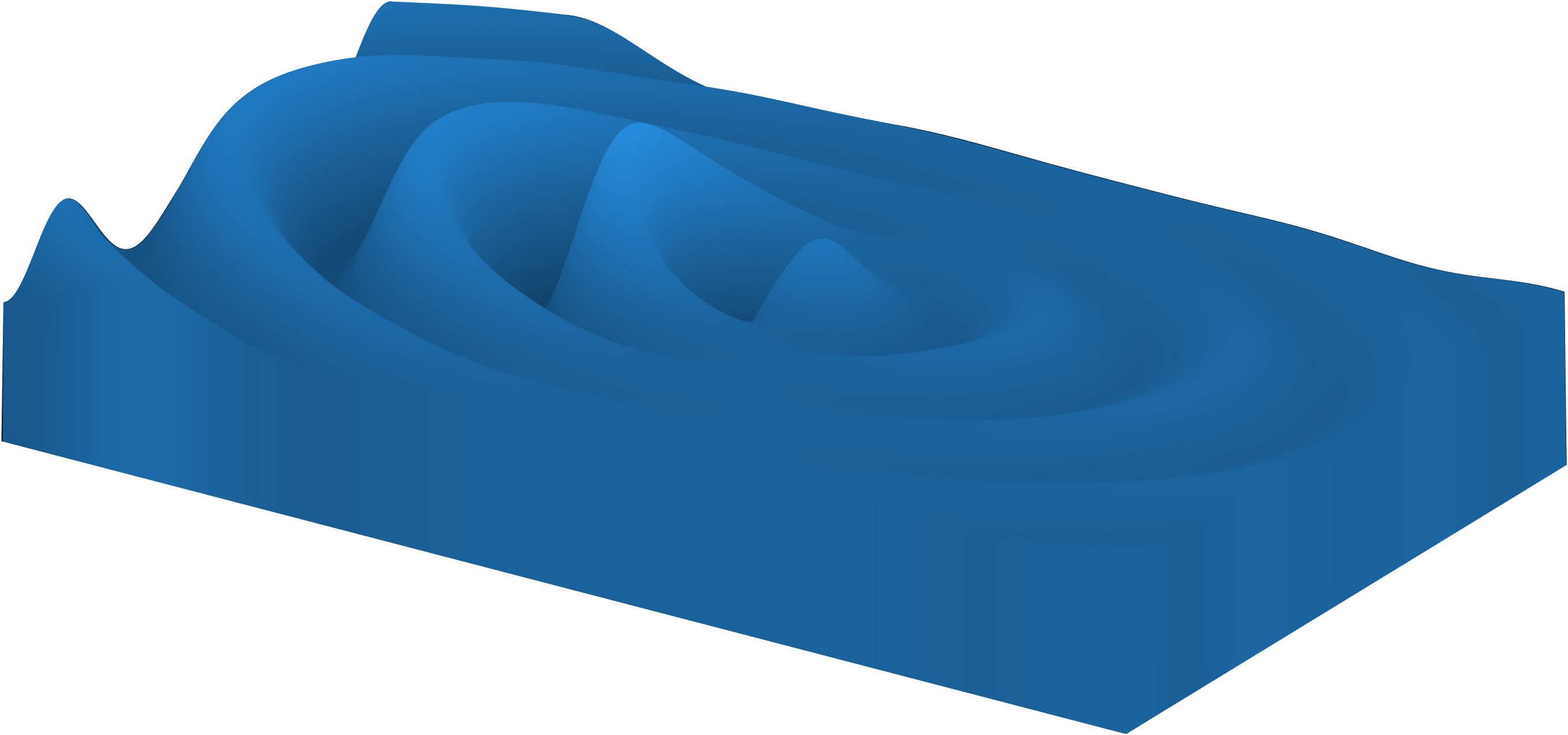}};

    \fill[rotate=-10,thick,red] (0.5,0.5) ellipse (1cm and 0.35cm);
		\draw[rotate=-10,thick] (0.5,0.5) ellipse (1cm and 0.35cm);
		\node[white] at (0.5,0.4) {$\mathcal{D}$};
		\node[white] at (1.5,-0.1) {$\partial\mathcal{D}$};
		\node[white] at (1.5,0.9) {$P(x,y,t)$};
		\draw[ultra thick,->] (-2,-2.4) -- (-2,-1.5);
		\draw[ultra thick,->] (-2,-2.4) -- (-1,-2);
		\draw[ultra thick,->] (-2.01,-2.4) -- (-1,-2.65);
		\node at (-2.3,-2) {$\tilde{z}$};
		\node at (-1.4,-1.8) {$\tilde{y}$};
		\node at (-1.5,-2.8) {$\tilde{x}$};
		\draw[<->,ultra thick,red] (-5.3,-0.5) -- (-5.3,0.4);
		\node[red] at (-5.6,-0.05) {$H$};
		\draw[->,ultra thick] (-4.65,0) -- (-4.65,0.2);
		\draw[->,ultra thick] (-4.65,1.3) -- (-4.65,1.1);
		\node[white] at (-4, 0.6) {$h(x,y,t)$};
		\draw[->,ultra thick,red] (1.5,2.15) -- (3,1.75);
		\node[red] at (2.4,2.4) {$U$};
		\draw (-5.04,0.42) -- (2.2,-1.47);
		\draw (-5.05,-0.48) -- (2.2,-2.37);
		\draw (2.2,-1.47) -- (2.2,-2.37);
		\draw (5.05,-0.63) -- (2.2,-2.37);
		\draw (5.05,-0.63) -- (5.03,0.27);
		\draw (5.03,0.27) -- (2.2,-1.47);
	\end{tikzpicture}
	\caption{A schematic of the two-dimensional shallow water set-up (where $k_{d}H\ll1$) where $k_{d}$ is the dimensional wavenumber and $H$ is the depth. A pressure source occupying the area $\mathcal{D}$ bounded by $\partial\mathcal{D}$ produces a wave field $h(x,y,t)$ where a fore-aft asymmetry in $\tilde{x}$ generates a net thrust in the positive $\tilde{x}$-direction driving a forward velocity $U$. The wave steepness is exaggerated for illustration purposes.}
	\label{fig:1}
\end{figure}
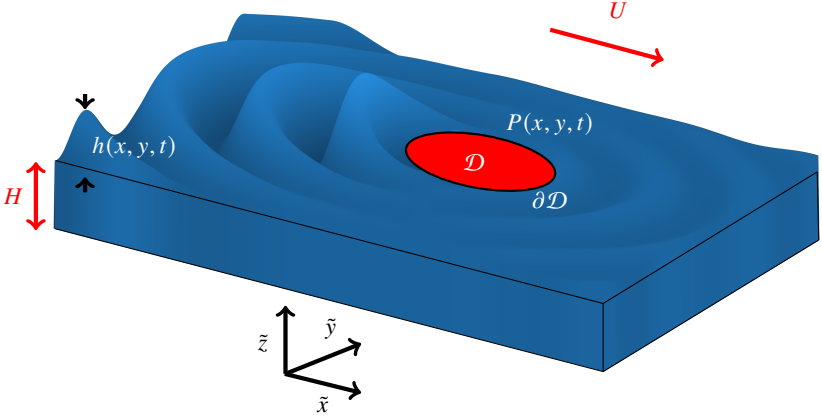

Consider a two-dimensional shallow water set-up where there is a pressure source $P(x,y,t)$ (representing a floating oscillating body) occupying the area $\mathcal{D}$ that is applied to the fluid surface (figure \ref{fig:1}). The definition of shallow water ($k_dH\ll1$, where $k_d$ is the dimensional wave number and $H$ is the depth of the fluid) is used to linearise the shallow water wave equations \citep{odonovan2026,book_shallow_water2006}.
We arrive at the following equation for the wave field $h(x,y,t)$:
\begin{align}\label{eq:dimensional_wave_equation}
	\nabla^2 h(x,y,t) -\frac{1}{c^2}\frac{\partial^2 h(x,y,t)}{\partial t^2} = -Q(x,y,t), && \mathrm{for }\,x,y\in(-\infty,\infty),
\end{align}
where $c=\sqrt{gH}$ is the wave speed, $g$ is acceleration due to gravity, $Q = \nabla^{2}P/\rho g$, with $\rho$ being density, and $\nabla^2 = \partial^2/\partial x^2 + \partial^2/\partial y^2$. We work in a non-dimensional set-up by scaling the variables with the oscillation frequency $\omega$ and wave speed $c$, similar to \citet{odonovan2026},
\begin{align}
	h,x,y\sim \frac{c}{\omega}, && t\sim\frac{1}{\omega}, && Q\sim\frac{\omega}{c}.
\end{align}
Therefore, the dimensionless wave equation is:
\begin{align}\label{eq:dimensionless_wave_equation}
	\nabla^2 h(x,y,t) -\frac{\partial^2 h(x,y,t)}{\partial t^2} = -Q(x,y,t), && \mathrm{for }\,x,y\in(-\infty,\infty).
\end{align}
The pressure source is used to model a raft at the fluid surface, which is assumed to move at a constant velocity $U$ in the $x$-direction. 
We will operate in the moving frame with dimensionless velocity $v=U/c$ using a Galilean transformation:
\begin{align}\label{eq:galilean_transformation}
	\tilde{x}=x-vt, && \tilde{y}=y, && \tilde{t} = t.
\end{align}
This yields the modified wave equation 
\begin{equation}\label{eq:wave_eq_moving_frame}
	(1-v^2)\frac{\partial^2h}{\partial \tilde{x}^2}(\tilde{x},\tilde{y},\tilde{t}) + 2v\frac{\partial^2h}{\partial \tilde{x}\,\partial\tilde{t}}(\tilde{x},\tilde{y},\tilde{t}) + \frac{\partial^2 h}{\partial \tilde{y}^2}(\tilde{x},\tilde{y},\tilde{t}) -\frac{\partial^2 h}{\partial \tilde{t}^2}(\tilde{x},\tilde{y},\tilde{t}) = -Q(\tilde{x},\tilde{y},\tilde{t}).
\end{equation}
The source is centred at the moving origin and defined over a finite region $\mathcal{D}$:
\begin{equation}\label{eq:piecewise_Q}
	Q(\tilde{x},\tilde{y},\tilde{t}) = 
	\begin{cases}
		Q(\tilde{x},\tilde{y},\tilde{t}), & \mathrm{for}\,\, (\tilde{x},\tilde{y})\in \mathcal{D}, \forall \tilde{t},
		\\
		0, & \mathrm{for}\,\, (\tilde{x},\tilde{y})\notin \mathcal{D}, \forall \tilde{t}.
	\end{cases}
\end{equation}
We assume that the source and wave field oscillate periodically in time:
\begin{align}\label{eq:periodic_assumption}
	Q(\tilde{x},\tilde{y},\tilde{t}) = \mathrm{Re}\left(\hat{Q}(\tilde{x},\tilde{y})\mathrm{e}^{-i\tilde{t}}\right), &&
	h(\tilde{x},\tilde{y},\tilde{t}) = \mathrm{Re}\left(\hat{h}(\tilde{x},\tilde{y})\mathrm{e}^{-i\tilde{t}}\right),
\end{align}
which simplifies the PDE further:
\begin{equation}\label{eq:convected_helmholtz}
	(1-v^2)\frac{\partial^{2}\hat{h}}{\partial \tilde{x}^2}(\tilde{x},\tilde{y}) -2iv\frac{\partial \hat{h}}{\partial\tilde{x}}(\tilde{x},\tilde{y}) +\frac{\partial^2\hat{h}}{\partial\tilde{y}^2}(\tilde{x},\tilde{y}) +\hat{h}(\tilde{x},\tilde{y}) = -\hat{Q}(\tilde{x},\tilde{y}).
\end{equation}
In Section \ref{sec3:startup}, we consider $v=0$ (the start-up case) resulting in the inhomogeneous Helmholtz equation. Here, we apply a radially outward Sommerfeld boundary condition in the far-field \citep{Devauchelle2020walkermemory}:
\begin{align}\label{eq:startup_bc_in_sec_2}
	\cos(\theta) \hat{h}_{\tilde{x}} + \sin(\theta) \hat{h}_{\tilde{y}} = i\hat{h},
	&&
	\textrm{as}\,\,\, \tilde{x},\tilde{y}\to\infty,
\end{align}
where $\theta$ is the angle measured from the positive $\tilde{x}$-axis and the subscript denotes partial differentiation.
We note that the periodic part of \eqref{eq:periodic_assumption} has a $-i\tilde{t}$ to stay consistent with the boundary conditions from \citet{Devauchelle2020walkermemory} rather than a $+i\tilde{t}$ convention in our previous work \citep{odonovan2026}. With $v\neq0$, \eqref{eq:convected_helmholtz} is the convected Helmholtz equation \citep{barucq2022_boundary_conds,hu2019_prandtl,becache2005_PML,marchner2021_PML_using_lorentz} and the boundary conditions imposed depend on whether the body is moving at a subcritical ($v<1$) or supercritical ($v>1$) velocity. Therefore,
the boundary conditions will be discussed separately for these cases in Sections \ref{sec4:subcrit} and \ref{sec6:supercrit}, respectively. We refer to the velocity $v=1$ as the critical case. As the body tends towards this velocity, the waves become increasingly steep which violate the linear assumption. Therefore, we ignore this case. 

Inspired by the derivation of the one-dimensional shallow water thrust \citep{odonovan2026}, we multiply \eqref{eq:dimensionless_wave_equation} by $\widetilde{\nabla}h$ and propose the dimensionless thrust per unit length is written as a vector decomposed into $\tilde{x},\tilde{y}$ components:
\begin{equation}\label{eq:2d_thrust_pde_system}
	\left((1-v^2)\frac{\partial^2h}{\partial \tilde{x}^2} + 2v\frac{\partial^2h}{\partial \tilde{x}\,\partial\tilde{t}} + \frac{\partial^2 h}{\partial \tilde{y}^2} -\frac{\partial^2 h}{\partial \tilde{t}^2}\right)\begin{pmatrix} h_{\tilde{x}} \\ h_{\tilde{y}}\end{pmatrix} = -Q\begin{pmatrix} h_{\tilde{x}} \\ h_{\tilde{y}}\end{pmatrix}.
\end{equation}
The first vector element is interpreted as an equality between the thrust injected (RHS) and the thrust radiated in the $\tilde{x}$-direction (LHS). In Appendix \ref{ap_sec:xthrust}, integration over the surface and a period of oscillation defines the time-averaged dimensionless thrust in the $\tilde{x}$-direction:
\begin{equation}\label{eq:thrust_general_velocity_expression}
	\begin{aligned}
	\bar{F}_{T} = \langle\hat{Q},\hat{h}_{\tilde{x}}\rangle= -\frac{1}{4}\int^{\tilde{y}^{+}}_{\tilde{y}^{-}}\bigg( 
		\left[|\hat{h}|^2 + 
		(1-v^2)|\hat{h}_{\tilde{x}}|^2
		- |\hat{h}_{\tilde{y}}|^2
		\right]_{\tilde{x}^{-}}^{\tilde{x}^{+}}   
	\bigg)\,\mathrm{d}\tilde{y}
	\\ 
	- \frac{1}{4}\int^{\tilde{x}^{+}}_{\tilde{x}^{-}}\bigg[\hat{h}_{\tilde{y}}^{*}\hat{h}_{\tilde{x}}+\hat{h}_{\tilde{y}}\hat{h}_{\tilde{x}}^{*}\bigg]^{\tilde{y}^{+}}_{\tilde{y}^{-}}\,\mathrm{d}\tilde{x},
	\end{aligned}
\end{equation}
where $\tilde{x}^{\pm},\tilde{y}^{\pm}$ denote the edges of a square region traced in the far-field away from the body, and $\langle\cdot,\cdot\rangle$ is defined as follows:
\begin{equation}
	\langle\hat{Q},\hat{h}_{\tilde{x}}\rangle = 
	\frac{1}{4}\int^{\tilde{x}^{+}}_{\tilde{x}^{-}}\int^{\tilde{y}^{+}}_{\tilde{y}^{-}} 
	\Big(\hat{Q}\hat{h}_{\tilde{x}}^{*} + \hat{Q}^{*}\hat{h}_{\tilde{x}}\Big)
	\,\mathrm{d}\tilde{x}\,\mathrm{d}\tilde{y}.
\end{equation}
The dimensionless thrust \eqref{eq:thrust_general_velocity_expression}  only differs from its dimensional counterpart by a $\rho gHc/\omega$ prefactor. 
This general expression is independent of the boundary conditions applied. Hence, it is defined over all velocity regimes. 
The first integral contains the familiar difference in fore--aft wave amplitude squared indicative of using the excess momentum flux of the waves for propulsion \citep{LONGUETHIGGINS1964,Longuet-Higgins1977meanforces}. There are two more terms in the first integral which correct the thrust due to the wave having both flux in the $\tilde{x}$- and $\tilde{y}$-directions. The first of these is seen in the one-dimensional case while the latter appears due to the introduction of $\tilde{y}$-dynamics. Meanwhile, the second integral accounts for the outflow of wave momentum from the sides of the body.

To aid interpretation, we prescribe a plane wave $\hat{h}=\mathrm{exp}(i\boldsymbol{k}\cdot\boldsymbol{\tilde{x}})$ to demonstrate the allowed wavenumbers geometrically for a given velocity (see Appendix \ref{ap_sec:xthrust}). With the two-dimensional wavenumber decomposed into $\tilde{x},\tilde{y}$ components $\boldsymbol{k}=(k_{(\tilde{x})},k_{(\tilde{y})})^{T}$, with the bracketed subscript denoting the $\tilde{x},\tilde{y}$ components rather than partial differentiation. The dispersion relation is a circle when $v=0$ and becomes an ellipse for $0<v<1$ before forming a hyperbola as $v>1$. Both curves extending into the negative direction when $v>1$ demonstrates the Doppler shift of the front wave to a rearwards travelling direction for supercritical velocities. In the absence of $\tilde{y}$-dynamics, the thrust reduces to that of \citet{odonovan2026} using the allowed wavenumbers. Hence, we define time-averaged thrust in the $\tilde{x}$-direction as $\bar{F}_{T}=\langle\hat{Q},\hat{h}_{\tilde{x}}\rangle$.

The thrust in the $\tilde{y}$-direction is interpreted as the second vector element of \eqref{eq:2d_thrust_pde_system}. It is shown in Appendix \ref{ap_sec:ythrust} that once an integral is taken over the surface and time-averaged, the thrust is defined as:
\begin{equation}\label{eq:ythrust}
\begin{aligned}
	\langle\hat{Q},\hat{h}_{\tilde{y}}\rangle=&
	-\frac{1}{4}\int^{\tilde{x}^{+}}_{\tilde{x}^{-}}
	\bigg(\left[|\hat{h}|^2 + |\hat{h}_{\tilde{y}}|^2
	- (1-v^2)|\hat{h}_{\tilde{x}}|^2\right]_{\tilde{y}^{-}}^{\tilde{y}^{+}}\bigg)\,\mathrm{d}\tilde{x}
	\\
	&- \frac{1-v^2}{4}\int^{\tilde{y}^{+}}_{\tilde{y}^{-}}\bigg[\hat{h}_{\tilde{y}}^{*}\hat{h}_{\tilde{x}}+\hat{h}_{\tilde{y}}\hat{h}_{\tilde{x}}^{*}\bigg]^{\tilde{x}^{+}}_{\tilde{x}^{-}}\,\mathrm{d}\tilde{y} 
	\\
	&
	+\frac{iv}{2}\Bigg(
		\int^{\tilde{y}^{+}}_{\tilde{y}^{-}}\bigg[\hat{h}\hat{h}_{\tilde{y}}\bigg]^{\tilde{x}^{+}}_{\tilde{x}^{-}} \,\mathrm{d}\tilde{y} + \int^{\tilde{x}^{+}}_{\tilde{x}^{-}}\bigg[\hat{h}\hat{h}_{\tilde{x}}\bigg]^{\tilde{y}^{+}}_{\tilde{y}^{-}} \,\mathrm{d}\tilde{x}
	\Bigg).
\end{aligned} 
\end{equation}
The coordinates are chosen for the body to move in the $\tilde{x}$-direction. Therefore, we expect the thrust in the $\tilde{y}$-direction to be zero. In Appendix \ref{ap_sec:ythrust}, we show that $\langle\hat{Q},\hat{h}_{\tilde{y}}\rangle=0$ when there is a symmetry in the wave field perpendicular to the direction of travel, as observed by \citet{Roh2019bee}. In the optimisation problem to follow, we will maximise thrust in the $\tilde{x}$-direction (henceforth referred to as just the thrust), and due to symmetry, the perpendicular thrust will be zero. 

Given a body with thrust $\bar{F}_{T}$, we can study the efficiency of propulsion:
\begin{equation}\label{eq:efficiency}
	\eta = \frac{\bar{F}_{T}}{\overline{\mathrm{Pow}}},
\end{equation}
where $\overline{\mathrm{Pow}}$ is the time-averaged power. In dimensional terms, this would be $c\cdot\bar{F}_{T}/\overline{\mathrm{Pow}}$, a ratio between power used in forward thrust divided by the total applied power.
To find the time-averaged power, we begin with the total energy in the moving frame,
\begin{equation}\label{eq:energy_moving_frame}
	E_{v}(\tilde{t}) = \frac{1}{2}\int^{\tilde{x}^{+}}_{\tilde{x}^{-}}\int^{\tilde{y}^{+}}_{\tilde{y}^{-}}
	\bigg[ (1+v^2)h_{\tilde{x}}^{2} - 2vh_{\tilde{x}}h_{\tilde{t}} + h_{\tilde{t}}^2 + h_{\tilde{y}}^2\bigg]\,\mathrm{d}\tilde{x}\,\mathrm{d}\tilde{y},
\end{equation}
and take a derivative with respect to time. Following the steps in Appendix \ref{ap_sec:power}, we find an equivalence relation between the power injected by the source and the power radiated away by the waves:
\begin{equation}
\begin{aligned}
	\overline{\mathrm{Pow}} = \langle\hat{Q},-i\hat{h}\rangle -v\langle\hat{Q},\hat{h}_{\tilde{x}}\rangle = 
	-\frac{1-v^2}{4}\int^{\tilde{y}^{+}}_{\tilde{y}^{-}}\Big[i\hat{h}_{\tilde{x}}\hat{h}^{*} - i\hat{h}_{\tilde{x}}^{*}\hat{h} \Big]^{\tilde{x}^{+}}_{\tilde{x}^{-}}\,\mathrm{d}\tilde{y} 
	\\
	-\frac{1}{4}\int^{\tilde{x}^{+}}_{\tilde{x}^{-}}\Big[i\hat{h}_{\tilde{y}}\hat{h}^{*} - i\hat{h}_{\tilde{y}}^{*}\hat{h} \Big]^{\tilde{y}^{+}}_{\tilde{y}^{-}}
	-v\bar{F}_{T}.
\end{aligned}
\end{equation}
The plane wave example is used again to understand the expression further in Appendix \ref{ap_sec:power}. It transpires that the $\tilde{y}$-direction has the sum of the right and left wave amplitudes squared, whilst the $\tilde{x}$-direction is harder to interpret due to the Doppler shift.
However, in the absence of $\tilde{y}$-dynamics, the one-dimensional power expression \citep{odonovan2026} returns.

The work to follow is concerned with optimising the thrust due to WDP \eqref{eq:thrust_general_velocity_expression} where the wave equation and boundary conditions act as constraints. A final constraint will be applied to regularise the problem. This will take the form of bounding the norm of the control function 
\begin{equation}\label{eq:norm}
	\iint_{\mathcal{D}} \hat{Q}^{*}\hat{Q}\,\mathrm{d}\tilde{x}\,\mathrm{d}\tilde{y}\leq\beta,
\end{equation}
where $\beta$ is a dimensionless bound set to $\beta=1$ for simplicity. The parameter $\beta$ in effect controls the magnitude of $\hat{Q}$, so increasing this value increases the effect of the source producing larger waves and thrust. This bound will also force the control function to be well-behaved with increased effect for smaller $\beta$ or larger areas. 

In our previous work \citep{odonovan2026}, we took a second case where the power was bounded and efficiency was optimised. Here, we will focus purely on the bounded norm case (maximising thrust), 
beginning with the start-up case ($v=0$) where we validate the numerical optimisation methods analytically in Section \ref{sec3:startup}.

\section{Start-up formulation and optimisation}\label{sec3:startup}

We begin with the start-up case, where the body is at rest and starting up to move forward (in the positive $\tilde{x}$-direction). Taking $v\approx0$, \eqref{eq:convected_helmholtz} becomes the inhomogeneous Helmholtz equation,
\begin{equation}\label{eq:startup_pde_helmholtz}
	\frac{\partial^2\hat{h}}{\partial x^2}(x,y) + \frac{\partial^2\hat{h}}{\partial y^2}(x,y) +\hat{h}(x,y) = -\hat{Q}(x,y). 
\end{equation}
The tilde is dropped from the coordinate system to signify $\tilde{x},\tilde{y}=x,y$ in the start-up case. With the body centred at the origin, we apply radial Sommerfeld boundary conditions in the far-field, similar to \citet{Devauchelle2020walkermemory}, to prevent reflection at the boundaries, 
\begin{align}\label{eq:startup_bc}
	&& \cos(\theta) \hat{h}_{x} + \sin(\theta) \hat{h}_{y} = i\hat{h}, && \mathrm{as}\,\,x,y\to\infty, &&
\end{align}
where $\theta$ is the angle measured from the positive $x$-axis. 
The boundary conditions are satisfied by the following Green's function solution
\begin{equation}\label{eq:startup_greens}
	\hat{h}(x,y) = \frac{i}{4}\iint_{\mathcal{D}}\hat{Q}(X,Y)H^{(1)}_{0}\left(|\boldsymbol{x}-\boldsymbol{X}| \right)\,\mathrm{d}X\,\mathrm{d}Y,
\end{equation}
where $H^{(1)}_{0}$ is the zeroth order Hankel function of the first kind \citep{Devauchelle2020walkermemory}. In this section, we define $\mathcal{D}$ in \eqref{eq:piecewise_Q} to be a square source of side length $l=L\omega/c$, where $L$ is the dimensional side length.

To pose the numerical optimisation, we truncate the infinite domain with the Sommerfeld boundary conditions \eqref{eq:startup_bc} applied at the edge of a domain five times larger than the body to operate in the far-field. 
The domain is discretised, approximating the derivatives with finite differences and the integrals with the trapezoid rule. The set-up forms algebraic constraints in the numerical optimisation implemented with JuMP in Julia \citep{Lubin2023jump} using the IPOPT solver \citep{wachter2006_IPOPT}.

\begin{figure}
	\centering
	\begin{tikzpicture}
		\node at (-4,0) {\includegraphics[width=0.5\textwidth]{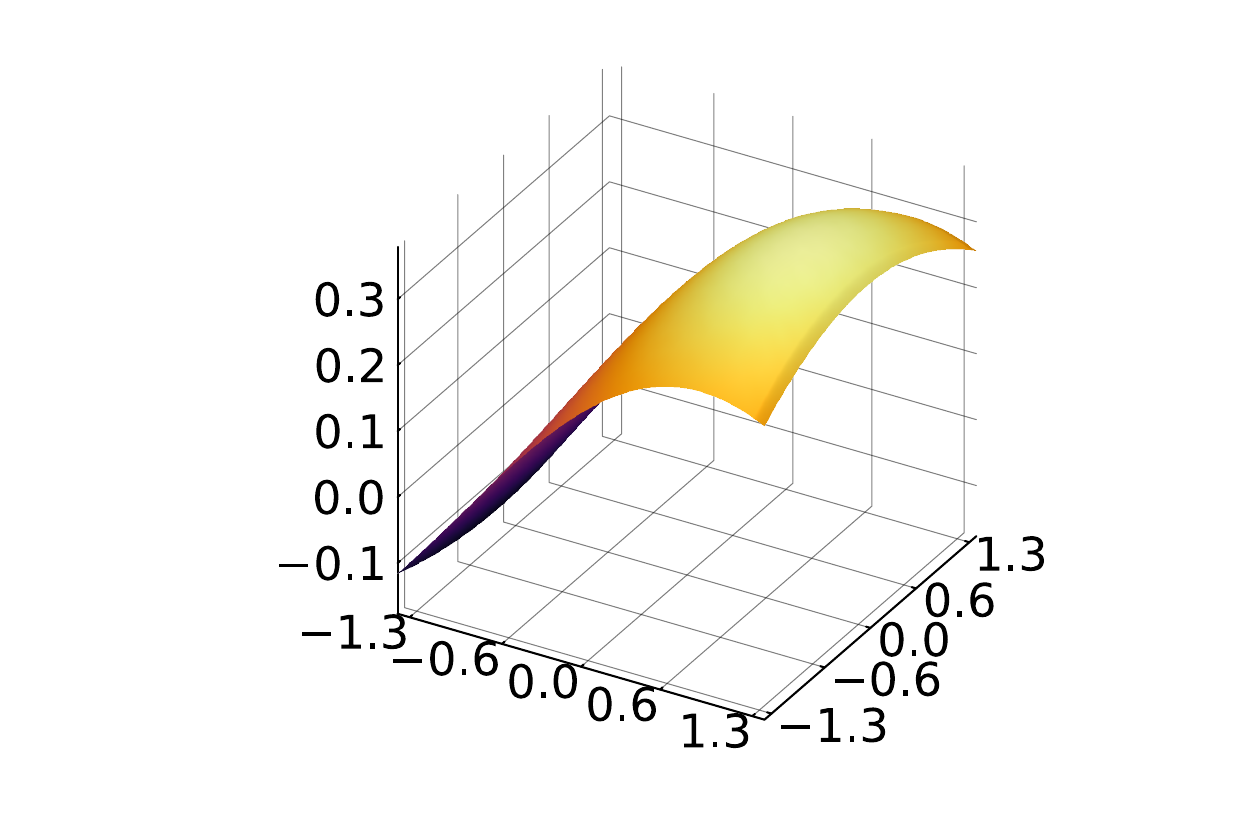}};
		\node at (-4,-4.5) {\includegraphics[width=0.5\textwidth]{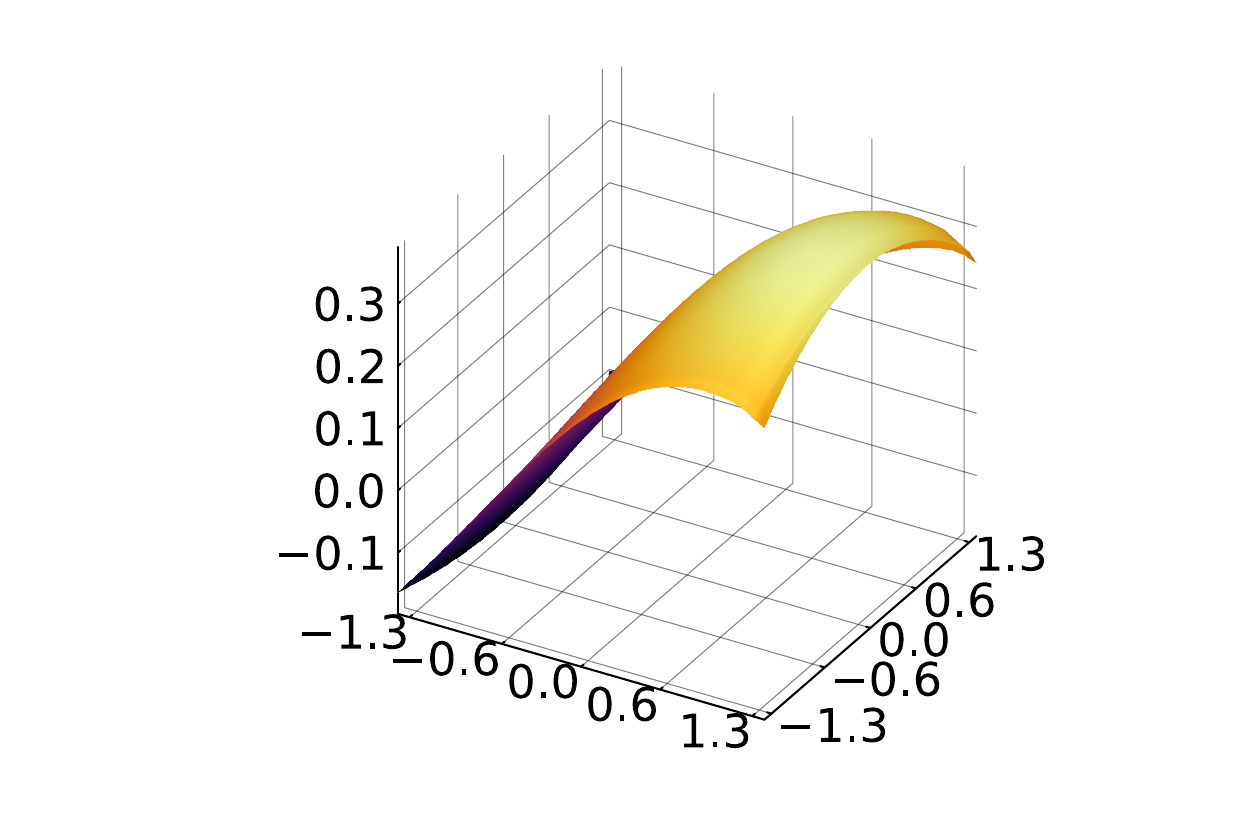}};
		\node at (2,0) {\includegraphics[width=0.5\textwidth]{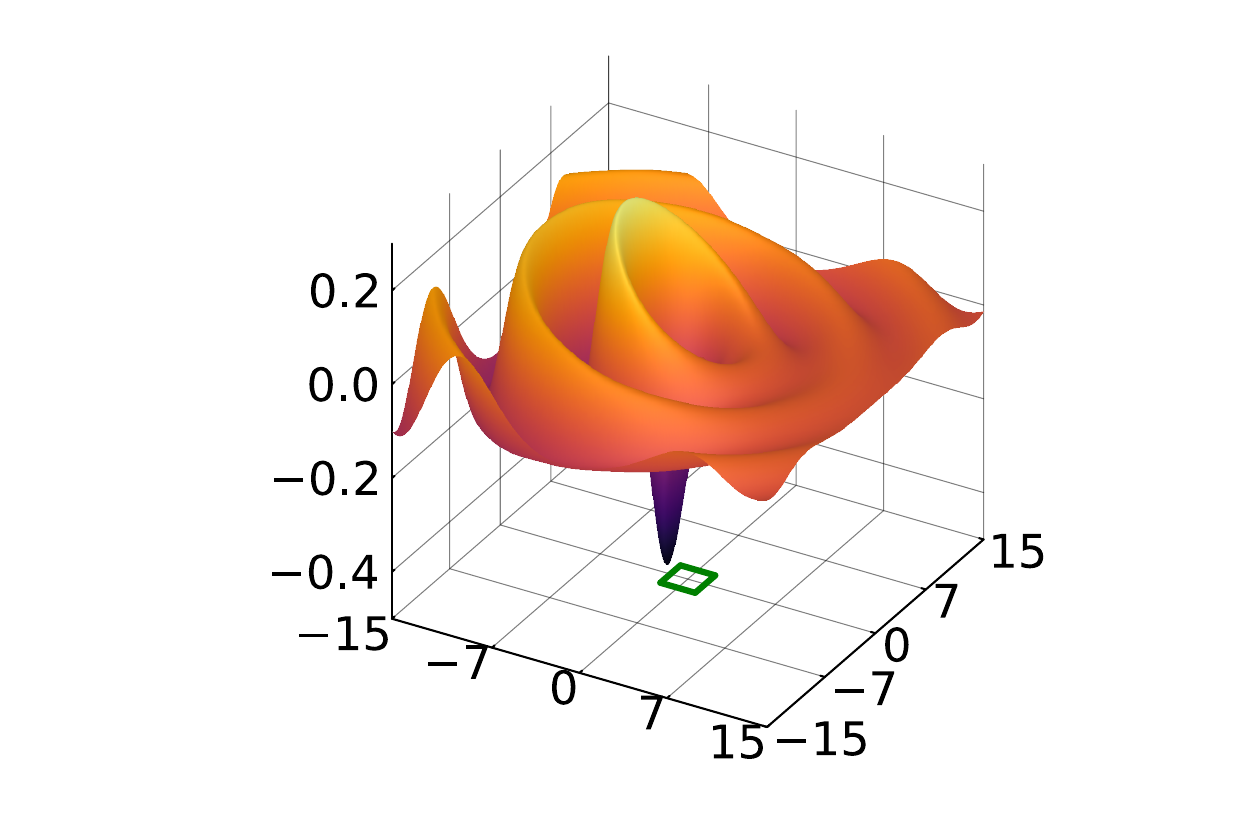}};
		\node at (1.8,-4.5) {\includegraphics[width=0.5\textwidth]{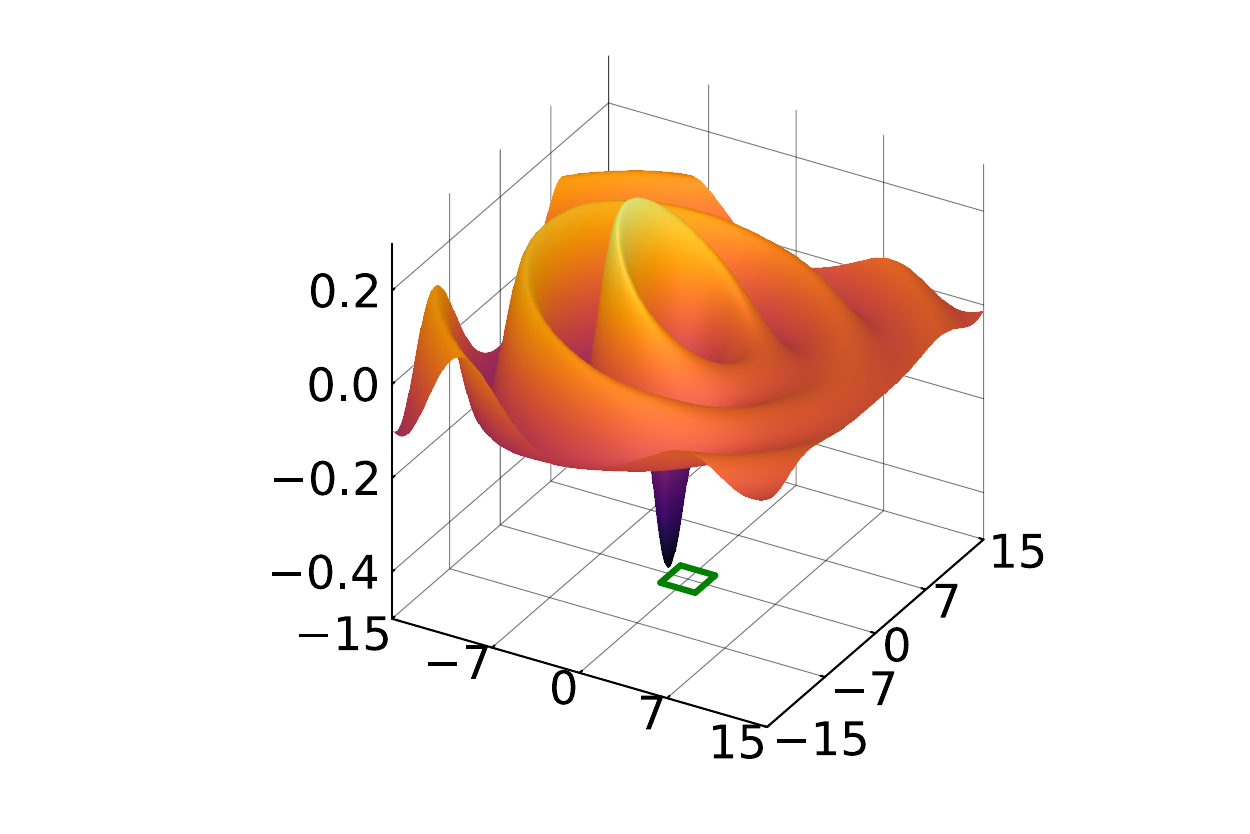}};
		\node at (-4.8,-1.8) {{$\tilde{x}$}};
		\node at (1.2,-1.8) {{$\tilde{x}$}};
		\node at (-4.8,-6.3) {{$\tilde{x}$}};
		\node at (1.2,-6.3) {{$\tilde{x}$}};
		\node at (-2,-1.5) {{$\tilde{y}$}};
		\node at (3.8,-1.5) {{$\tilde{y}$}};
		\node at (-2,-6) {{$\tilde{y}$}};
		\node at (3.8,-6) {{$\tilde{y}$}};
		\node at (-6.5,2) {{$(a)$}};
		\node at (-1,2) {{$(b)$}};
		\node at (-6.5,-2.5) {{(c)}};
		\node at (-1,-2.5) {{(d)}};
		\node [rotate=90] at (-6.5,0) {$\mathrm{Re}(\hat{Q})$};
		\node [rotate=90] at (-6.5,-4.5) {$\mathrm{Re}(\hat{Q})$};
		\node [rotate=90] at (-0.5,0) {$\mathrm{Re}(\hat{h})$};
		\node [rotate=90] at (-0.5,-4.5) {$\mathrm{Re}(\hat{h})$};
	\end{tikzpicture}
	
	\caption{(a) The optimal square source for $v=0$ and $l=\pi^{3/2}/2$ in a $150\times150$ grid. (b) The corresponding wave field to the source in (a) whose side length is represented by the green square for scale. (c) The optimal source generated under the same conditions through the analytically informed optimisation. It appears similar, but varies slightly due to the discretisation of the governing equation. (d) The wave field produced by the source in (c).}
	\label{fig:analytical opti}
\end{figure}
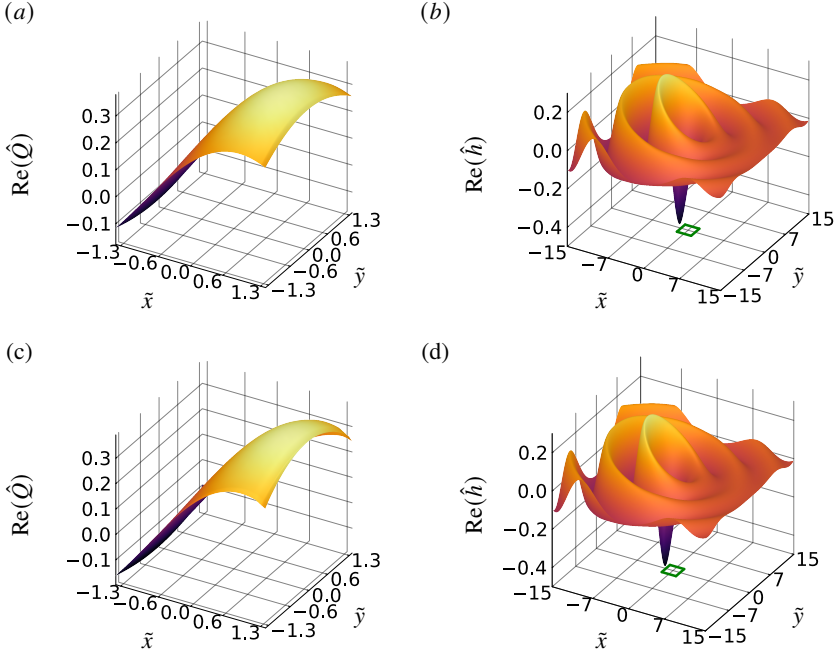

The resulting optimal source for $l=\pi^{3/2}/2$ is shown in figure \ref{fig:analytical opti}$a$. The expected difference in fore--aft amplitude along with a symmetry in the $\tilde{y}$-direction is seen in figure \ref{fig:analytical opti}$b$. In this set-up, the source cannot eliminate the side waves completely. The waves in the $\tilde{y}$-direction consume some of the injected power (whilst not contributing to the thrust) which prevents $\eta=1$ being achieved. Instead, $\eta\approx0.67$ is achieved in the bounded norm case.

To validate the optimisation methods, we turn to analytical optimisation with variational calculus.
Substituting the wave field \eqref{eq:startup_greens} into the thrust expression \eqref{eq:thrust_general_velocity_expression}, we aim to minimise the functional
\begin{equation}\label{eq:v0_norm_functional}
	f = \bar{F}_{T} + 2\lambda \langle\hat{Q},\hat{Q}\rangle,
\end{equation}
where $\lambda$ is a Lagrange multiplier. By perturbing the functional (see Appendix \ref{ap_sec:start-up_norm}), the optimal condition on $\hat{Q}$ is found to be 
\begin{equation}\label{eq:norm_perturb_result}
	-\frac{i}{8} \int^{\frac{l}{2}}_{-\frac{l}{2}}\int^{\frac{l}{2}}_{-\frac{l}{2}}\hat{Q}(X,Y)\frac{x-X}{|\boldsymbol{x}-\boldsymbol{X}|}J_{1}(|\boldsymbol{x}-\boldsymbol{X}|)\,\mathrm{d}X\,\mathrm{d}Y + \lambda \hat{Q}(x,y) = 0. 
\end{equation}
The first term is in the null space of the two-dimensional Helmholtz operator $\mathcal{L}_{2D} = \partial_{xx}  +\partial_{yy}+\mathbb{I}$. Therefore, by applying $\mathcal{L}_{2D}$ to \eqref{eq:norm_perturb_result}, the optimal condition is:
\begin{equation}\label{eq:v0_norm_LQ}
	\mathcal{L}_{2D}\hat{Q}=0.
\end{equation}
This is the same result as the one-dimensional case \citep{odonovan2026}. Indeed, both the real and imaginary part of $\hat{Q}$ from the numerical optimisation satisfy \eqref{eq:v0_norm_LQ} to the order $\sim10^{-8}$, up to two numerical indices inward from the edge of the $\hat{Q}\neq0$ region (where the transition to $\hat{Q}=0$ creates errors). This validates our result, and the next obvious step is to solve \eqref{eq:v0_norm_LQ} to learn more about the solution. 

Given the apparent shape of the optimal source in figure \ref{fig:analytical opti}$a$, we postulate that \eqref{eq:v0_norm_LQ} can be solved with separation of variables in $x,y$, denoting the separation constant with $\kappa$. The resulting source is composed of four individual wave-like components, 
\begin{equation}\label{eq:startup_optimal_Q_wave_solutions}
	\hat{Q}(x,y) =  A \mathrm{e}^{i\kappa x+i\sqrt{1-\kappa^2}y} + B \mathrm{e}^{-i\kappa x+i\sqrt{1-\kappa^2}y} + C \mathrm{e}^{i\kappa x-i\sqrt{1-\kappa^2}y} + D \mathrm{e}^{-i\kappa x-i\sqrt{1-\kappa^2}y},
\end{equation}
which account for the contributions in the $\pm x$- and $\pm y$-directions. Rather than returning to \eqref{eq:norm_perturb_result} to solve the resulting integral equation for the complex constants $A$-$D$, we will take a simpler approach to validate. We instead enforce \eqref{eq:startup_optimal_Q_wave_solutions} as a constraint to find the constants $A$-$D$ and $\kappa$ via numerical optimisation and compare the solutions. This will be referred to as the analytically informed optimisation, and the resulting source is displayed in figure \ref{fig:analytical opti}$c$.

Qualitatively, the solutions look close to each other. 
We take a quantitative measure of the correspondence between the two solutions by finding the largest absolute difference. This value is calculated to be at maximum $4.5\times10^{-2}$ for a $150\times150$ grid omitting the 2 indices from the boundary where $\hat{Q}$ varies more due to the discontinuity at the square edge. The solutions are close, but there is a small error due to difficulty in converging to the same optimum in a solution set with an arbitrary complex argument.  The first optimisation also uses a discretised (approximate) Helmholtz equation, while the second uses the exact solution to the Helmholtz equation. This is reflected in the fact that the analytically informed optimal solution does not satisfy the discretised $\mathcal{L}_{2D}\hat{Q}=0$ to the same order of error ($\sim10^{-4}$) as the pure numerical optimisation ($\sim10^{-8}$).
The constants $A,B,C,D,\kappa$ are extracted from the analytically informed optimisation and their values are displayed in table \ref{tab3:ana_opti_vals}. 
The constants control the shape of the source in $x$ and $y$. To create a symmetry in $y$ in the resulting wave field we will need $A=C$ which is approximately achieved as seen in table \ref{tab3:ana_opti_vals}. Similarly, in the rearward wave there is a symmetry in the $y$-direction with $B\approx D$. A fore--aft asymmetry is achieved through $A\neq B$ and $C\neq D$.

\begin{table}
	\centering
	\begin{tabular}{ccccc}
		$A$  & $B$ & $C$ & $D$ & $\kappa$ \\
		$-0.01218 + -0.01708i$ & $0.14529 +0.11994i$ & $-0.01412 -0.01688i$ & $0.15080 +0.11315i$ & $0.85856$ \\
	\end{tabular}
	\caption{Resulting values (to $5$ d.p.) for the constants in \eqref{eq:startup_optimal_Q_wave_solutions} from the analytically informed optimisation over a grid of $150\times150$ points (figure \ref{fig:analytical opti}$c$).}
	\label{tab3:ana_opti_vals}
\end{table}

The thrust from the analytically informed optimisation is $\bar{F}_{T}=0.2459$ with $\overline{\mathrm{Pow}}=0.3707$. Meanwhile, the purely numerical scheme outputs $\bar{F}_{T}=0.2489$ and $\overline{\mathrm{Pow}}=0.3729$. We attribute the respective $1.2\%$ and $0.5\%$ error to the difference between the discretised and the analytical solution discussed earlier. The relative accuracy in the thrust and power is closer than the difference in shape. The solutions are expected to be unique up to an arbitrary complex argument, so tuning the starting guess to converge to the same local maximum proves difficult. Therefore, a similar approach will be used by comparing the thrust output in the sections to follow.

\section{Modelling a source at subcritical velocities}\label{sec4:subcrit}

\begin{figure}
	\centering
	\begin{tikzpicture}
		\node at (-3.2,0) {\includegraphics[width=0.55\textwidth]{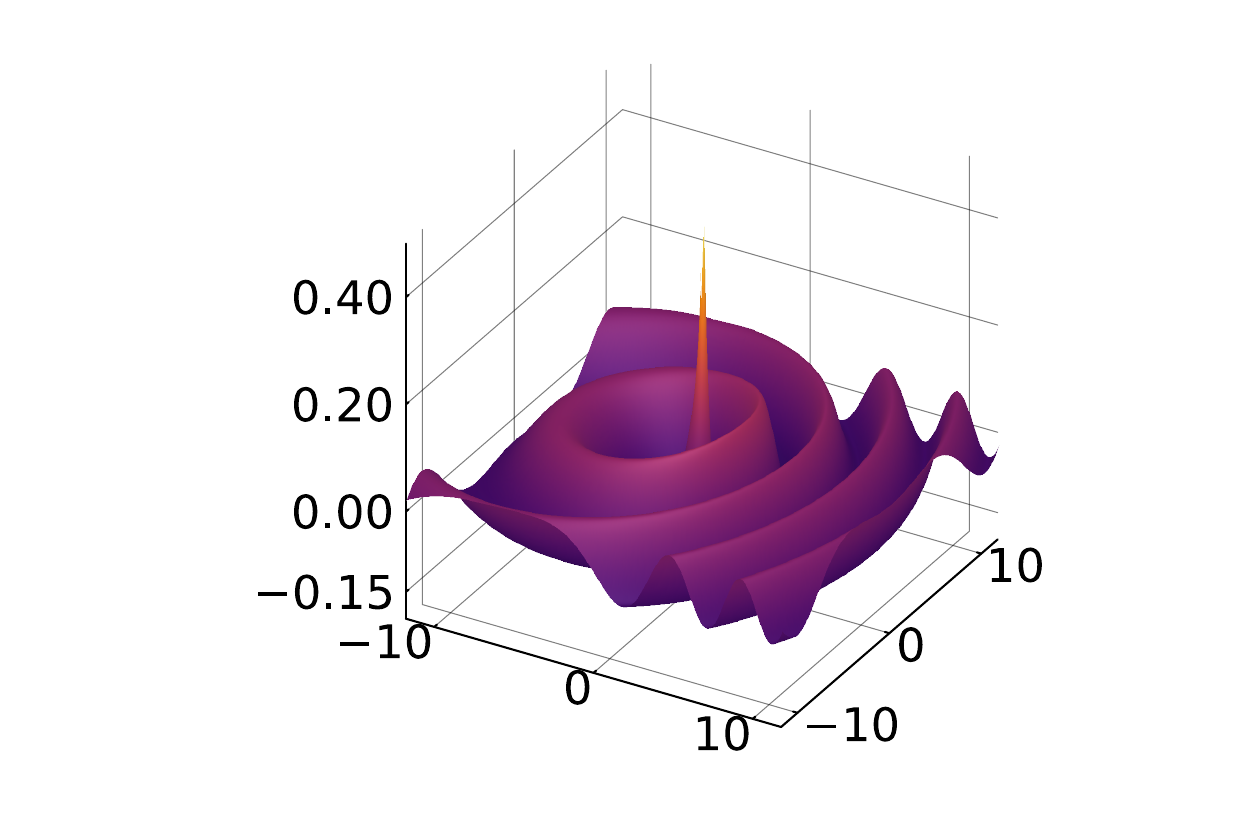}};
		\node at (3.2,0) {\includegraphics[width=0.55\textwidth]{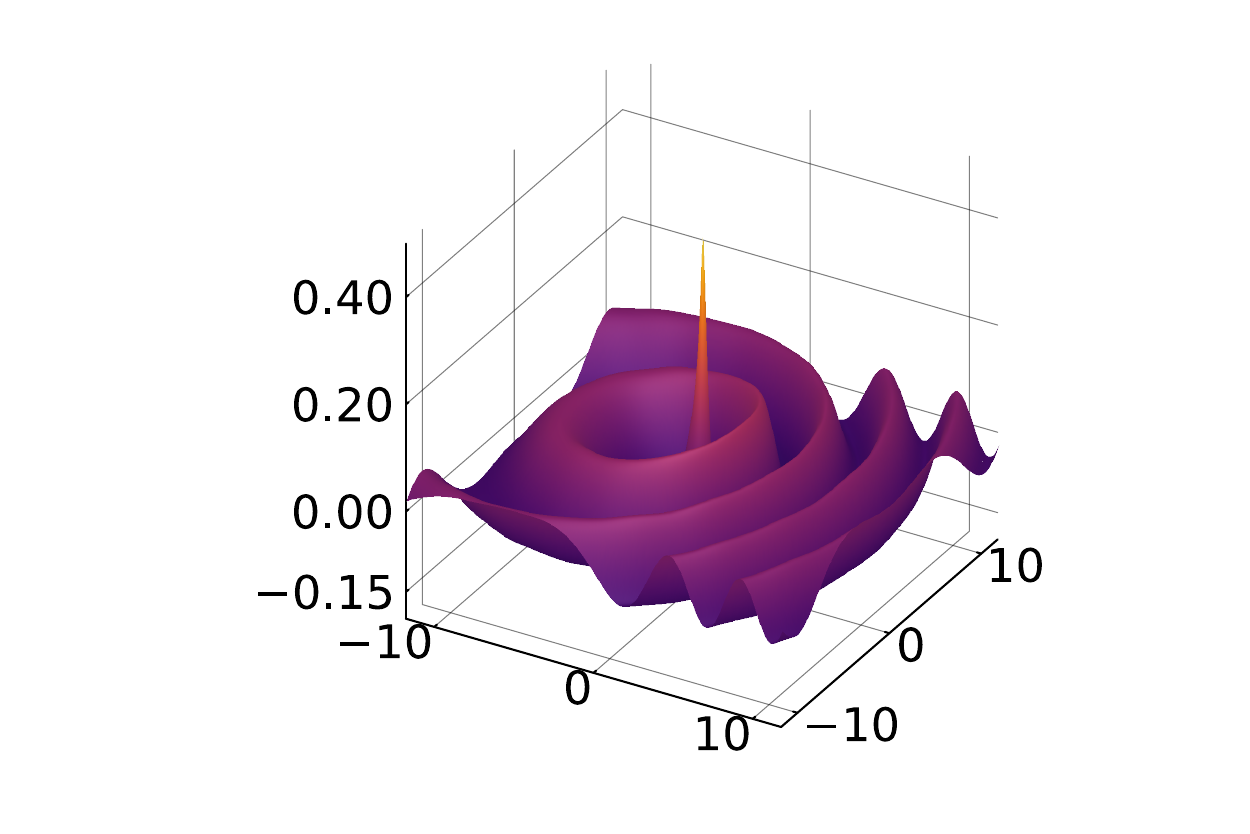}};
		\node at (1.2,2) {$(b)$};
		\node at (-5.2,2) {$(a)$};
		\node[rotate=90] at (-5.6,0) {$\mathrm{Re}(\hat{h})$};
		\node[rotate=90] at (0.8,0) {$\mathrm{Re}(\hat{h})$};
		\node at (-3.7,-2.05) {$\tilde{x}$};
		\node at (2.7,-2.05) {$\tilde{x}$};
		\node at (-1,-1.6) {$\tilde{y}$};
		\node at (5.4,-1.6) {$\tilde{y}$};
	\end{tikzpicture}
	\caption{(a) The wave field $\hat{h}$ due to a point source moving at $v=0.5$ found using the analytical solution \eqref{eq:gal_transform_pt_source_solution}. (b) The wave field found by solving \eqref{eq:convected_helmholtz} numerically and applying \eqref{eq:subcritical boundary condition} at the boundary.}
	\label{fig:hankel_numerical_subcritical_compare}
\end{figure}
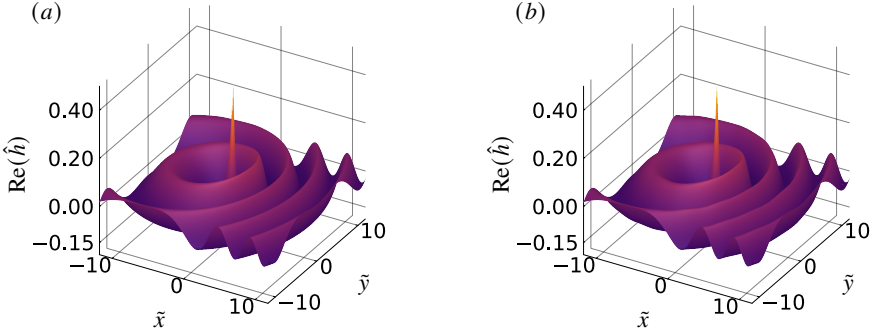

With the introduction of velocity, the radially outwards waves in the rest frame will undergo a Doppler shift in the moving frame of reference. Hence, the boundary condition \eqref{eq:startup_bc} is no longer valid. In one dimension, the upstream wave has a shorter wavelength while the downstream wave has a stretched wavelength in the moving frame. However, in two dimensions the Doppler shift is more complicated. This is highlighted when the dispersion relation for a plane wave source changes from a circle to an ellipse in Appendix \ref{ap_sec:xthrust}. 
To find the appropriate boundary condition, we will exploit the fact that the dimensional wave equation is invariant under a Lorentz transformation and its inverse:
\begin{equation}\label{eq:lorentz_transform}
	\begin{aligned}
		x = \gamma(x'+Ut'), && && x' =\gamma(x-Ut),
		\\
		y = y', && && y' =y,
		\\
		t = \gamma \left(t'+\frac{U}{c^2}x'\right), && && t' =\gamma\left(t-\frac{U}{c^2}x\right),
	\end{aligned}
\end{equation}
where $\gamma=1/\sqrt{1-v^2}$.
Using a point source, we show how \eqref{eq:dimensional_wave_equation} undergoes a Lorentz transformation in Appendix \ref{ap_sec:sub_bc}. Due to the invariance under a Lorentz transformation, the Green's function from \eqref{eq:startup_greens} still applies in a moving frame. We return the wave field to the rest frame before transforming into the Galilean moving frame. This results in the wave field produced by a moving point source in the coordinate system of interest ($\tilde{x},\tilde{y}$):
\begin{equation}\label{eq:gal_transform_pt_source_solution}
	\hat{h}(\tilde{x},\tilde{y}) = \frac{i\gamma}{4}H^{(1)}_{0}\left(\frac{\omega\gamma}{c}(\gamma^2\tilde{x}^2+\tilde{y}^2)^{\frac{1}{2}}\right)\mathrm{e}^{i\omega\gamma^2\frac{U}{c^2}\tilde{x}}.
\end{equation}
Motivated by the start-up boundary condition, we aim to find a far-field boundary condition of the form:
\begin{equation}\label{eq:general_form_bc_for_subbing_into}
		\boldsymbol{\hat{n}}\cdot\widetilde{\nabla} \hat{h} = \boldsymbol{\hat{n}}\cdot i\boldsymbol{k}\hat{h},
\end{equation}
for the normal vector $\boldsymbol{\hat{n}}$.
By deriving boundary conditions for a point source, we can impose these generally in the far-field, where any source approximates a point source.  
Far from the source, $\widetilde{\nabla}\hat{h}=(\hat{h}_{\tilde{x}},\hat{h}_{\tilde{y}})$ can be written in terms of $\hat{h}$ itself correct to first order (see Appendix \ref{ap_sec:sub_bc}). The unit normal vector $\boldsymbol{\hat{n}}$ is written in terms of the effective angle $\tilde{\theta}$, measuring the angle from the positive $\tilde{x}$-axis in the Galilean moving frame with a correction from $\gamma$:
\begin{equation}\label{eq:sub_normal_vec}
	\boldsymbol{\hat{n}} = 
	\begin{pmatrix}
		\cos(\tilde{\theta})
		\\
		\sin(\tilde{\theta})
	\end{pmatrix}
	= 
	\begin{pmatrix}
		\frac{\gamma\tilde{x}}{\sqrt{\gamma^2\tilde{x}^2+\tilde{y}^2}}
		\\
		\frac{\tilde{y}}{\sqrt{\gamma^2\tilde{x}^2+\tilde{y}^2}}
	\end{pmatrix}. 
\end{equation}
Hence, the far-field boundary condition for subcritical velocities in the dimensionless set-up is:
\begin{equation}\label{eq:subcritical boundary condition}
	\frac{1}{\gamma}\cos(\tilde{\theta})\hat{h}_{\tilde{x}} + \sin(\tilde{\theta})\hat{h}_{\tilde{y}} = 
	i\gamma\hat{h}\left(1 + v\cos(\tilde{\theta})\right),
	\,\,\,\,\tilde{x},\tilde{y}\to\infty.
\end{equation}
When $v=0$ is taken, \eqref{eq:subcritical boundary condition} simplifies to \eqref{eq:startup_bc}.

With the boundary condition applied to the numerical solver, we show that \eqref{eq:subcritical boundary condition} produces close to the same result as \eqref{eq:gal_transform_pt_source_solution} using a narrow Gaussian distribution to approximate the point source, as seen in  
figure \ref{fig:hankel_numerical_subcritical_compare}. For a more quantitative measure, we find the maximal absolute difference (omitting the centre since there is a singularity) is of the order $\sim10^{-3}$ for velocities $v<0.7$ and then increases to $\sim0.03$ by $v=0.8$. 
The numerical scheme sufficiently approximates the wave field with the chosen boundary conditions at lower subcritical velocities. However, it becomes less accurate as $v\to1$ because the wavelength of oscillation $\lambda_{\mathrm{w}}\to0$ in the upstream wave.

When $v\to1$, the forward travelling wavelength $\lambda_{\mathrm{w}}\to0$, so $k\gg1$, where $k$ is the magnitude of the corresponding two-dimensional wavenumber $\boldsymbol{k}=(k_{(\tilde{x})},k_{(\tilde{y})})$. However, the shallow water set-up requires that the condition $k_{d}H\ll1$ holds, where the subscript denotes the dimensional wavenumber. 
We will assess this condition using maximum wave steepness scaled with maximum wave amplitude, $k_{\mathrm{prox}}=\max|\widetilde{\nabla}\hat{h}|/\max|\hat{h}|$ as an appropriate proxy. With a wavenumber too large, the wave field would require a non-linear model to account for higher wave steepness.
Evaluating this on the wave field produced by a Gaussian source satisfying the norm constraint, we find that $k_{\mathrm{prox}}<1$ for $v<0.3$, and $k_{\mathrm{prox}}<5.5$ for $v<0.95$. This suggests our linear assumption may remain valid even up to $v\approx0.95$.
\begin{figure}
	\centering
	\begin{tikzpicture}
		\node at (3.0,0) {\includegraphics[width=0.4\textwidth]{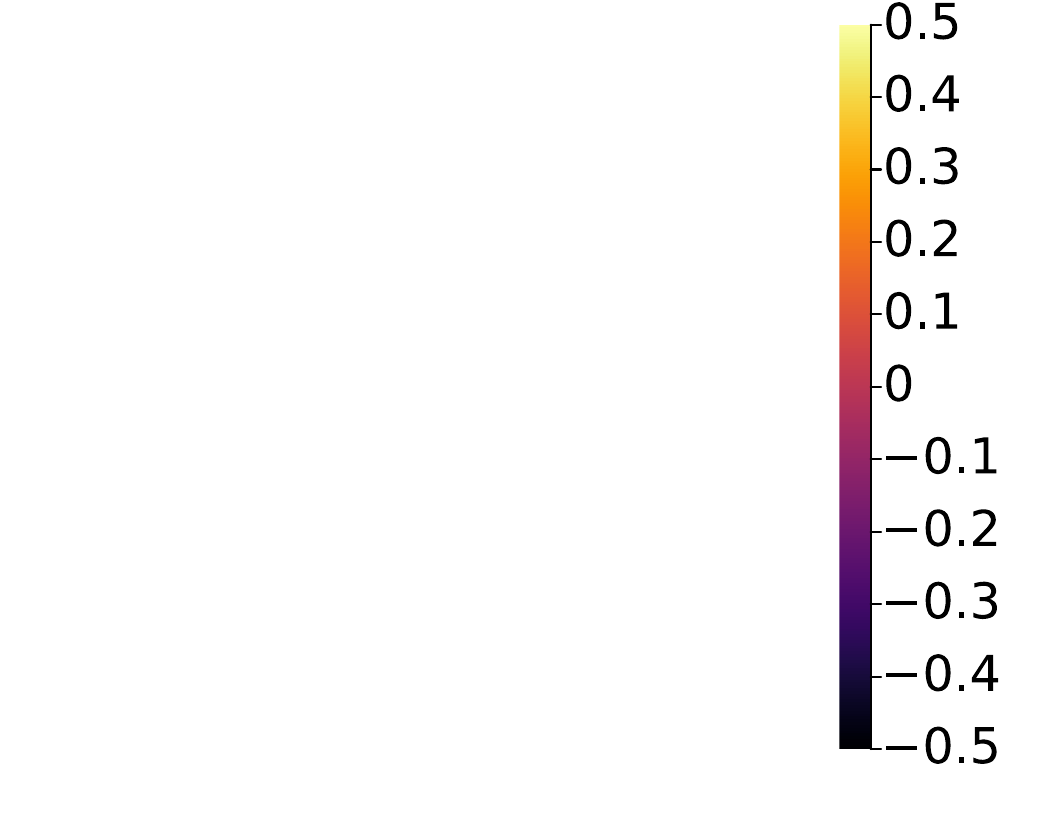}};
		\node at (-3.5,0) {\includegraphics[width=0.35\textwidth]{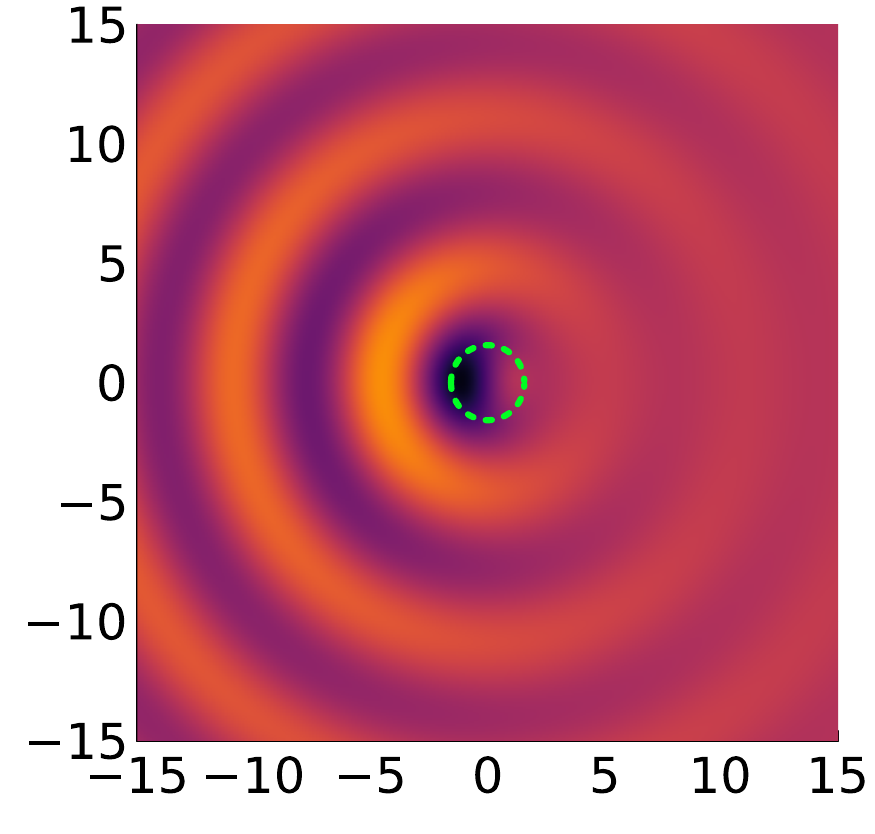}};
		\node at (2.3,0) {\includegraphics[width=0.35\textwidth]{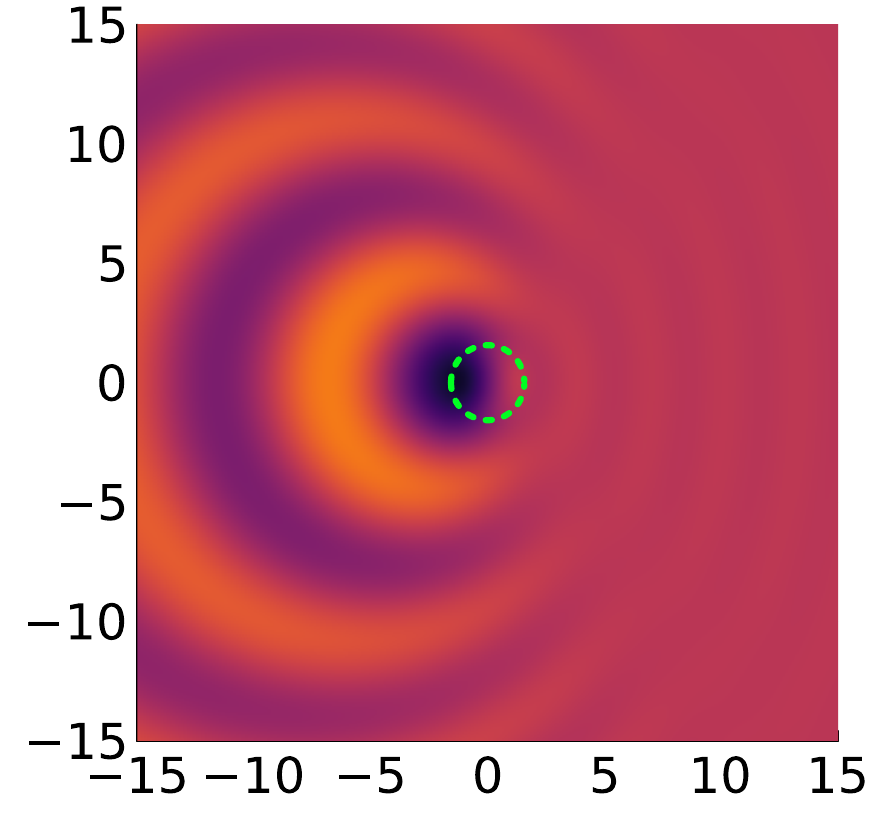}};
		\node at (-2,1.42) {\includegraphics[width=0.12\textwidth]{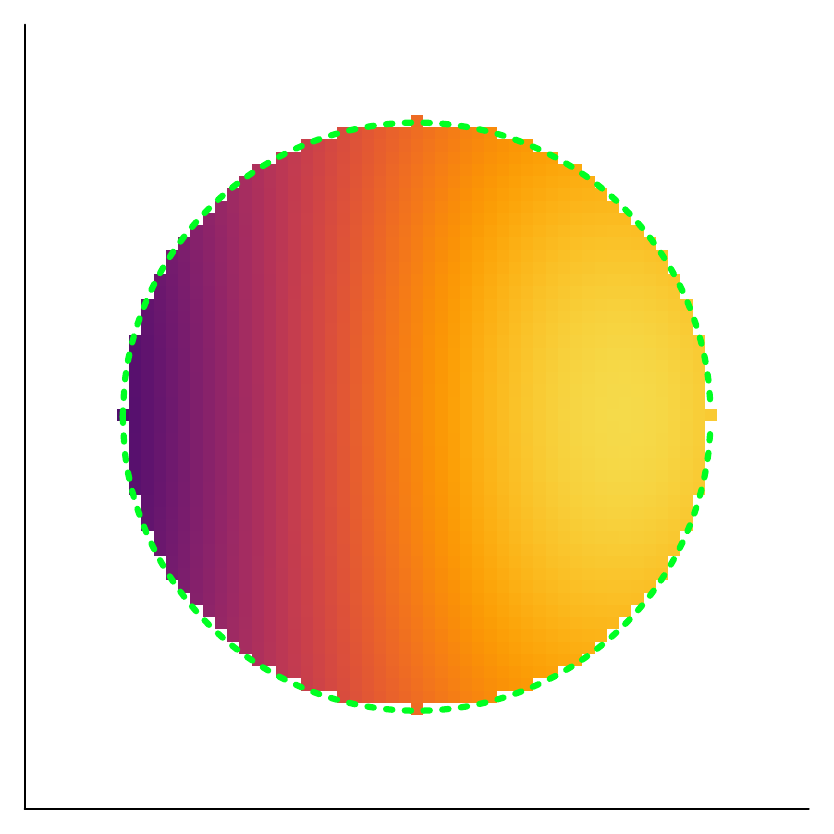}};
		\node at (3.8,1.42) {\includegraphics[width=0.12\textwidth]{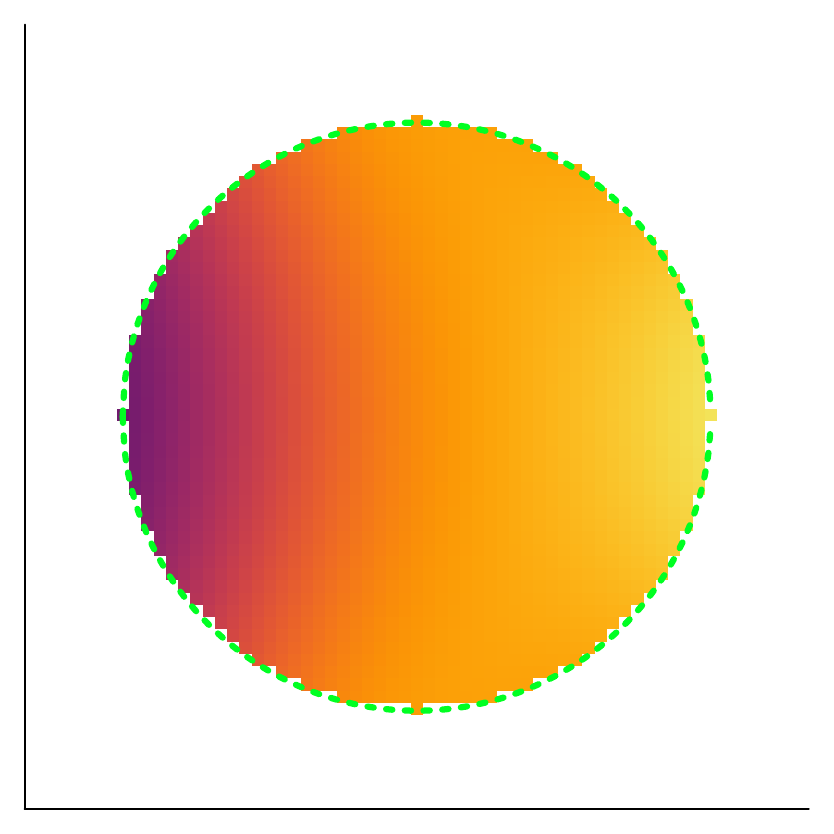}};
		\node at (-5.8,2.5) {$(a)$};
		\node at (0,2.5) {$(b)$};
		\node at (-3.2,-2.4) {$\tilde{x}$};
		\node at (2.6,-2.4) {$\tilde{x}$};
		\node at (-5.8,0.25) {$\tilde{y}$};
		\node at (0,0.25) {$\tilde{y}$};
		\node[rotate=90] at (5.8,0.3) {$\mathrm{Re}(\hat{h}),\mathrm{Re}(\hat{Q})$};
	\end{tikzpicture}
	\caption{(a) A top-down view of the real part of the optimal wave field for a circular body (radius $\pi/2$) with $v=0$. The real part of the source, whose location in the main plot is highlighted with the green circle, is found in the top corner. (b) A similar plot for $v=0.5$ where the colour bar corresponds both the wave field and source in both graphs.}
	\label{fig4:subcritical_numerical_result}
\end{figure}
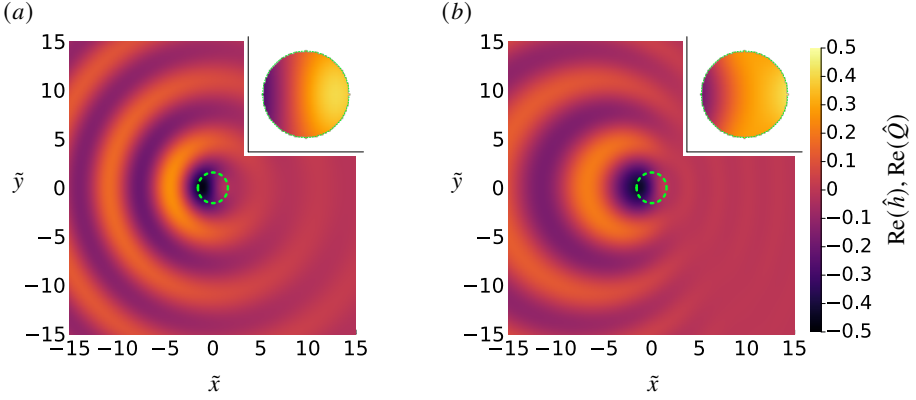

\subsection{Investigating the optimal source}\label{subsec4:subcritical_optimisation}

Similar to Section \ref{sec3:startup}, we next pose a numerical optimisation. The thrust is the objective while \eqref{eq:convected_helmholtz} and \eqref{eq:subcritical boundary condition} apply constraints. The norm of the control function \eqref{eq:norm} is once again bounded with $\beta=1$, but we will now use a circular source region $\mathcal{D}$ with radius $r=\pi/2$ to match the area of the square source. An example of the resulting optimal wave field is shown in figure \ref{fig4:subcritical_numerical_result} for velocities $v=0$ and $v=0.5$, where the former is presented for comparison to the square used in the start-up case. Comparing values, the square produces $\bar{F}_{T}=0.2489$ with $\eta=0.6708$ while the circle produces $\bar{F}_{T}=0.2276$ with $\eta=0.6465$. The square is marginally better, but we will proceed with the circle for the moment.

When $v=0$, we found that $\mathcal{L}_{2D}\hat{Q}=0$ holds. We can ask whether the same condition holds in the general subcritical case. Evaluating $|\mathcal{L}_{2D}\hat{Q}|$ numerically over velocities $v<1$, we find the maximum deviation from $\mathcal{L}_{2D}\hat{Q}=0$ to be of order $\sim10^{-6}$ without the noise at the boundary reported in the square case. This is still relatively small when compared with the source of order $\sim10^{-1}$. 
Thus, we claim that $\mathcal{L}_{2D}\hat{Q}=0$ is satisfied for subcritical velocities. 
To support this, we again apply separation of variables with separation constant $\kappa$. This results in the following $\hat{Q}$:
\begin{equation}
\begin{aligned}\label{eq:subandsup_sep_variables_result}
	\hat{Q}(\tilde{x},\tilde{y}) &=  A\exp\left(\frac{1}{2}
		\left[
			\frac{2iv}{1-v^2} 
			- \sqrt{-\frac{4v^2}{(1-v^2)^2}-\frac{4\kappa^2}{1-v^2}}
		\right]\tilde{x} + i\sqrt{1-\kappa^2}\,\tilde{y}\right)
	\\
	& +B\exp\left(\frac{1}{2}
		\left[
			\frac{2iv}{1-v^2} 
			- \sqrt{-\frac{4v^2}{(1-v^2)^2}-\frac{4\kappa^2}{1-v^2}}
		\right]\tilde{x} - i\sqrt{1-\kappa^2}\,\tilde{y}\right)
	\\
	& +C\exp\left(\frac{1}{2}
		\left[
			\frac{2iv}{1-v^2} 
			+ \sqrt{-\frac{4v^2}{(1-v^2)^2}-\frac{4\kappa^2}{1-v^2}}
		\right]\tilde{x} + i\sqrt{1-\kappa^2}\,\tilde{y}\right)
	\\
	& +D\exp\left(\frac{1}{2}
		\left[
			\frac{2iv}{1-v^2} 
			+ \sqrt{-\frac{4v^2}{(1-v^2)^2}-\frac{4\kappa^2}{1-v^2}}
		\right]\tilde{x} - i\sqrt{1-\kappa^2}\,\tilde{y}\right),
\end{aligned}
\end{equation}
where $A,B,C,D$ are complex constants.
An analytically informed subcritical optimisation is compared to the current optimisation thrust results. Over subcritical velocities with a source of radius $\pi/2$ the relative difference remains below $1\%$ for $v<0.4$ (the start-up circle performs better than a square) and is less than $8\%$ until $v=0.95$. This is the region where the gradients increase and the numerical scheme struggles, so the analytical solution not working as well here is to be expected.
We could go into more analytical detail from here (e.g. attempting to derive $\mathcal{L}_{2D}\hat{Q}=0$ using variational calculus), but this is not our focus. We must note that the analytically informed optimisation error reduces with a harsher norm constraint achieved by either reducing $\beta$ or increasing the area (e.g. error $<6\%$ for $r=3\pi/4,\beta=1$).

\subsection{A long or a wide source?}\label{subsec:sub_aspectratio}

\begin{figure}
	\centering
	\begin{tikzpicture}
		\node at (-3.3,0) {\includegraphics[width=0.45\textwidth]{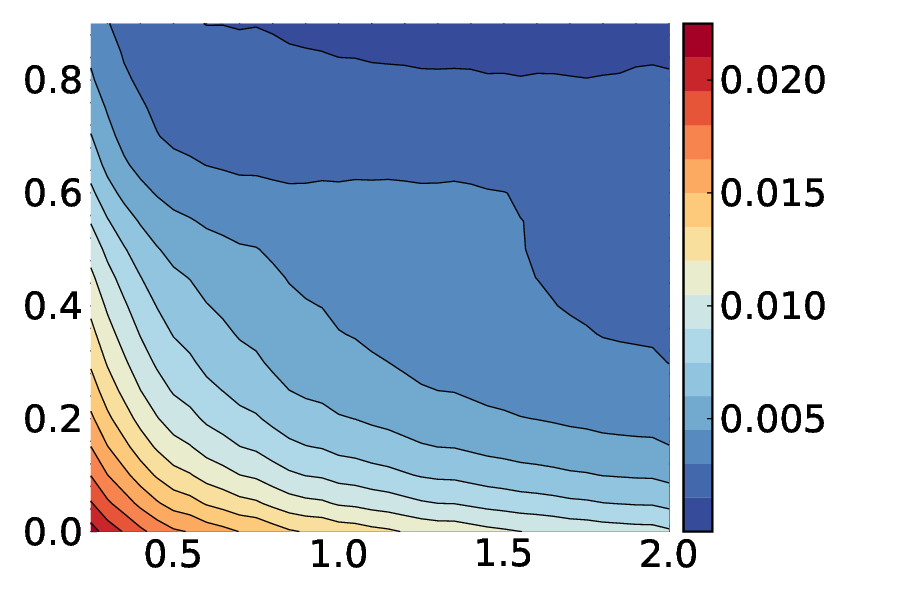}};
		\node at (3.5,0) {\includegraphics[width=0.45\textwidth]{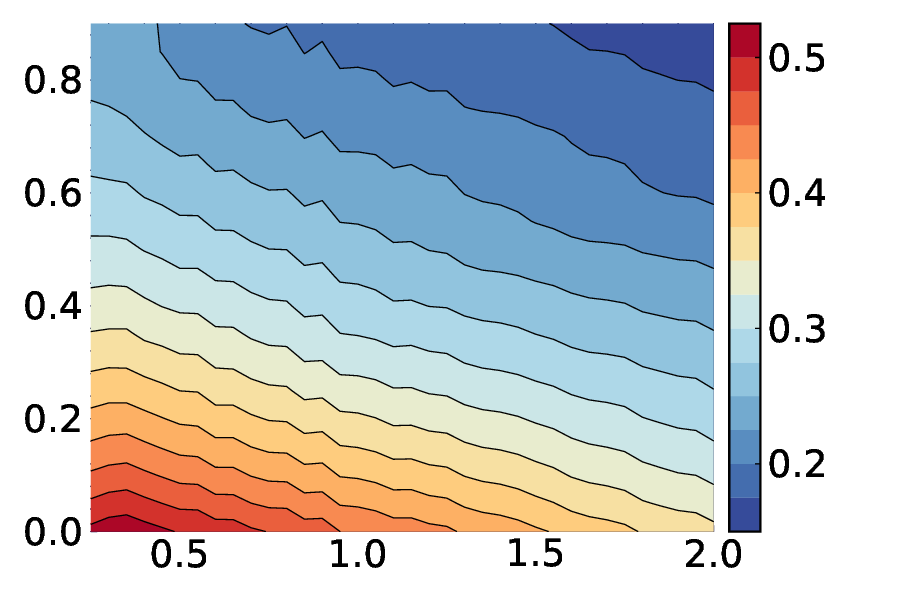}};
		\node[rotate=90] at (-6.4,0) {velocity ($v$)};
		\node[rotate=90] at (6.4,0) {$\bar{F}_{T}$};
		\node[rotate=90] at (0.4,0) {velocity ($v$)};
		\node[rotate=90] at (-0.4,0) {$\bar{F}_{T}$};
		\node at (-3.65,-2.1) {Aspect ratio ($b/a$)};
		\node at (3.15,-2.1) {Aspect ratio ($b/a$)};
		\node at (-6.4,2) {$(a)$};
		\node at (0.4,2) {$(b)$};
	\end{tikzpicture}
	\caption{(a) Contour plot to demonstrate the effect of varying velocity and aspect ratio $b/a$ on thrust for an elliptic source with constant area equivalent to a circle of radius $\pi/6$. (b) A similar contour plot where the source has constant area equivalent to a circle with radius $3\pi/4$. Note the varying colourbars. This is due to a smaller source being unable to produce as much thrust.}
	\label{fig:sub_sweeps}
\end{figure}
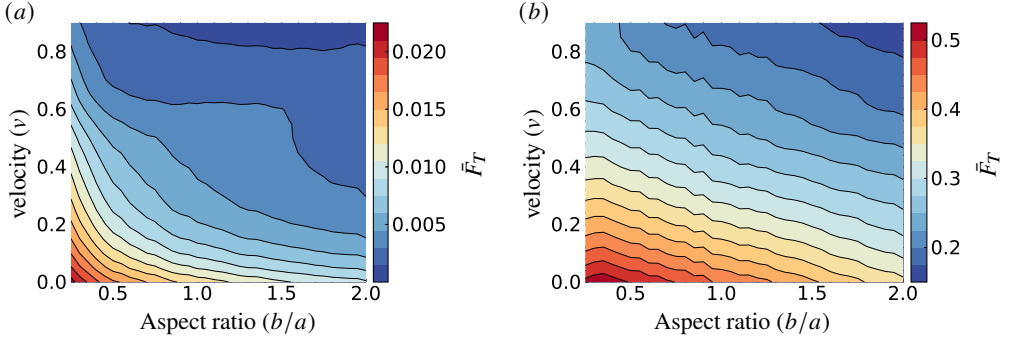

We want to understand how varying the aspect ratio affects maximal thrust.  
To define the aspect ratio, we can generalise to an elliptical source:
\begin{equation}\label{eq:elliptical_source}
	\frac{\tilde{x}^2}{a^2} + \frac{\tilde{y}^2}{b^2} = 1,
\end{equation}
where $a,b$ are the semi-major or semi-minor axes of the ellipse in $\tilde{x}$ and $\tilde{y}$, respectively. Defining the aspect ratio as $b/a$, we vary the shape of the source for a given velocity and evaluate thrust keeping area fixed. 
To study the effect of size, we will consider two raft areas, one larger and another smaller than the $r=\pi/2$ circle used already. 
In figure \ref{fig:sub_sweeps}$a,b$ we investigate ellipses of fixed areas $\pi a b =\pi^3/36$ and $9\pi^3/16$, respectively, demonstrating the case of both a small body and a large one.

A larger body produces a greater maximal thrust due to a larger integral domain. 
This is reflected in the differing force magnitude between the plots in figure \ref{fig:sub_sweeps}.
However, figure \ref{fig:sub_sweeps} primarily shows that thrust depends on aspect ratio.
In figure \ref{fig:sub_sweeps}$b$, a longer body ($a>b$) produces more thrust for a given velocity. For a fixed area, a longer body can accommodate more wavelengths allowing greater manipulation of the waves to generate forward thrust. A wider body ($b>a$) can accommodate fewer wavelengths in $\tilde{x}$, but the extra width over $\tilde{y}$ does not make up for this, resulting in a reduction in thrust. 
For a smaller source (figure \ref{fig:sub_sweeps}$a$) the preference for length ($a>b$) continues, albeit with a different pattern.

The results in figure \ref{fig:sub_sweeps} demonstrate the dependence of thrust on aspect ratio and velocity for a given area. We also see that thrust decreases with velocity, which is a common feature of such systems (e.g. \citep{odonovan2026}). 
We have so far looked at optimising the source and its aspect ratio in the case of a rectangle and an ellipse. We can also consider more general shapes by treating the boundary curve as a control to be optimised. 
This question will be considered in Section \ref{sec:optimal_shaping} once the supercritical case has been discussed. 

\section{Modelling a supercritical optimisation}\label{sec6:supercrit}

When $v>1$, the governing PDE \eqref{eq:convected_helmholtz} becomes hyperbolic. Therefore, the solution will evolve over characteristic trajectories. Transforming \eqref{eq:convected_helmholtz} into canonical form, we find the characteristics to be lines with slope $\pm\widehat{\gamma} = \pm1/\sqrt{v^2-1}$, labelled with reference to its close resemblance to the Lorentz factor $\gamma$:
\begin{align}\label{eq:sup_characteristics}
	\tilde{y} = -\widehat{\gamma}\tilde{x}+c_{0},
	&&
	\tilde{y} =\widehat{\gamma}\tilde{x}+d_{0},
\end{align}
where $c_{0}$ and $d_{0}$ are constants. Due to causality, the solution evolves outwards within an expanding wedge with angle ($\tan\theta =\widehat{\gamma}$), which may have a discontinuity at the boundary. Rather than the fixed wedge angle of the deep water Kelvin wake result \citep{rabaud2013_kelvin_or_mach}, this is a shallow water velocity-dependent wedge angle similar to a Mach wedge for large $v$.

In the supercritical case, the source is moving quicker than the waves it produces. Therefore, no wave can propagate upstream in the moving frame, which is imposed with the following boundary conditions:
\begin{align}
	&& && \hat{h}\left(\frac{a}{2},\tilde{y}\right) = \frac{\partial\hat{h}}{\partial\tilde{x}}\left(\frac{a}{2},\tilde{y}\right) = 0: && \tilde{y}\in(-\infty.\infty), && &&
\end{align}
where $a$ is the extent of the body in the positive $\tilde{x}$-direction following the convention of \eqref{eq:elliptical_source}.
The waves are only allowed to propagate within the wedge behind the source bounded by the characteristics \eqref{eq:sup_characteristics} originated from the front of the source. Therefore, the domain is chosen to be large enough in the $\tilde{y}$-direction to avoid any reflection off the boundaries.
A Crank-Nicolson scheme \citep{Crank_Nicolson_1947} is implemented into the same optimisation approach, in which the scheme marches backwards in $\tilde{x}$ behind the raft with a Cartesian grid with non-square grid spacing $dy>dx$. The numerical solutions have been validated against existing solvers in the \textit{SciML MethodOfLines.jl} package \citep{jones_2022_MOL}.

\begin{figure}
	\centering
	\begin{tikzpicture}
		\node at (-3.8,0) {\includegraphics[width=0.4\textwidth]{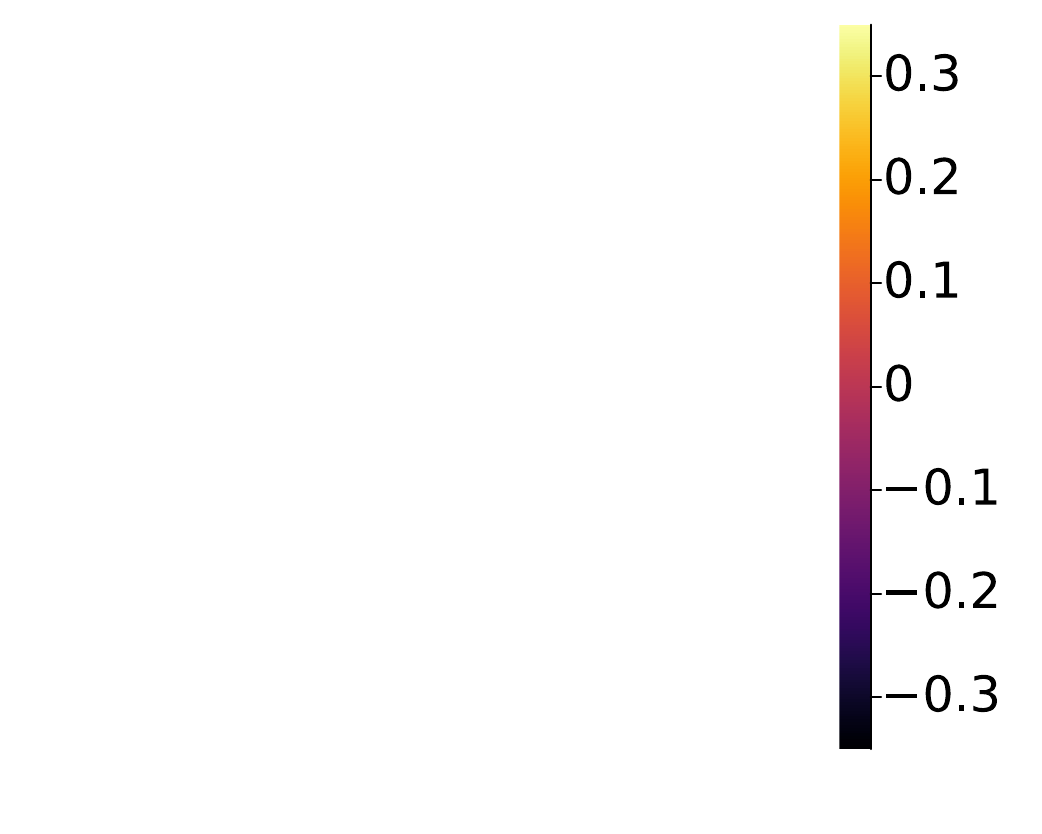}};
		\node at (3.2,1.3) {\includegraphics[width=0.35\textwidth]{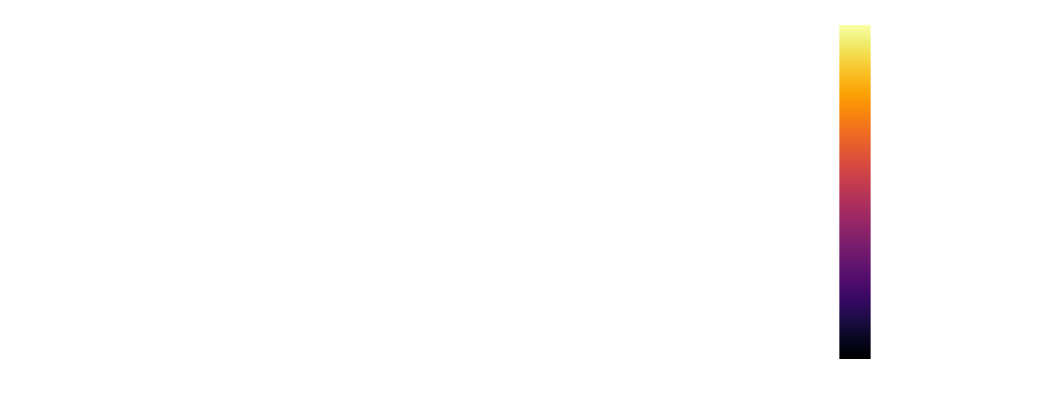}};
		\node at (-4.5,0) {\includegraphics[width=0.35\textwidth]{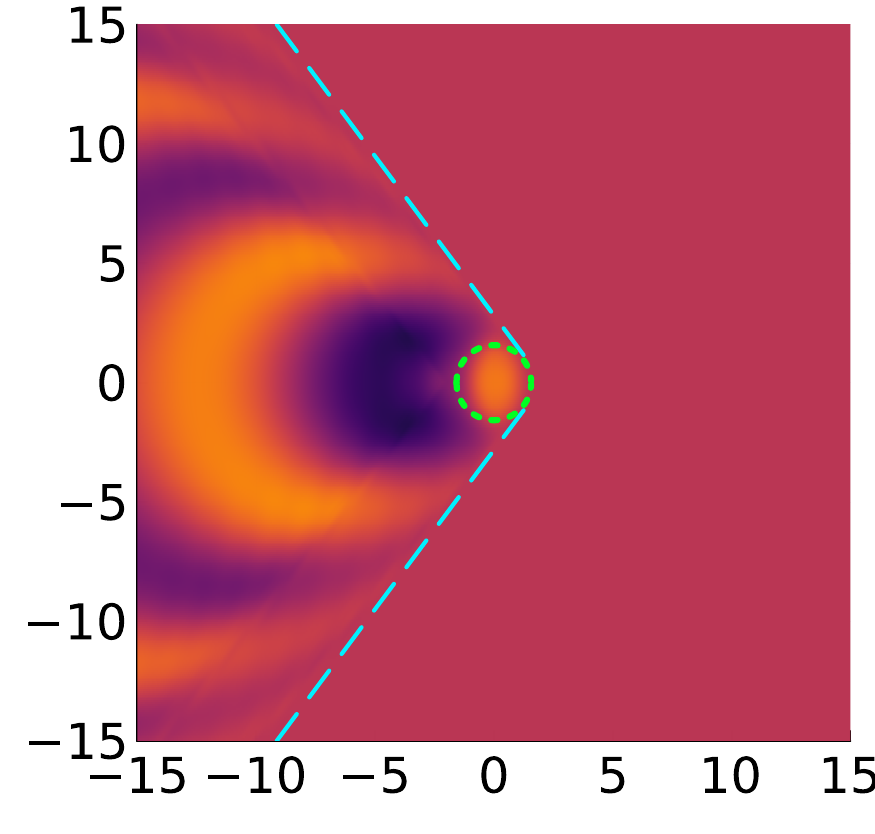}};
		\node at (2.3,0) {\includegraphics[width=0.35\textwidth]{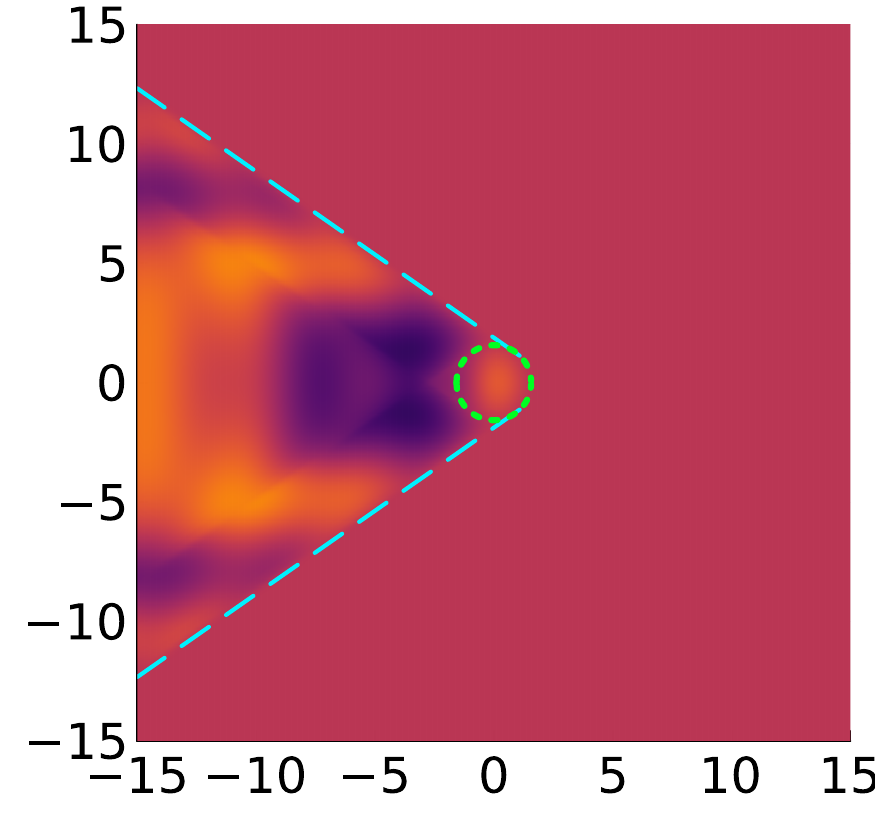}};
		\node at (-3.1,1.4) {\includegraphics[width=0.125\textwidth]{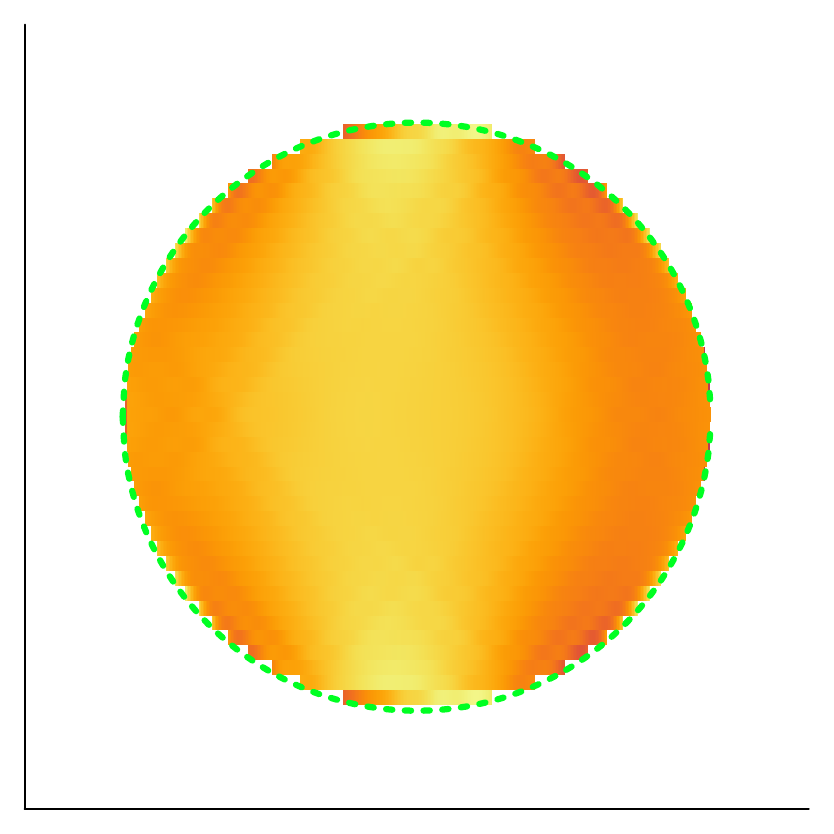}};
		\node at (3.7,1.4) {\includegraphics[width=0.125\textwidth]{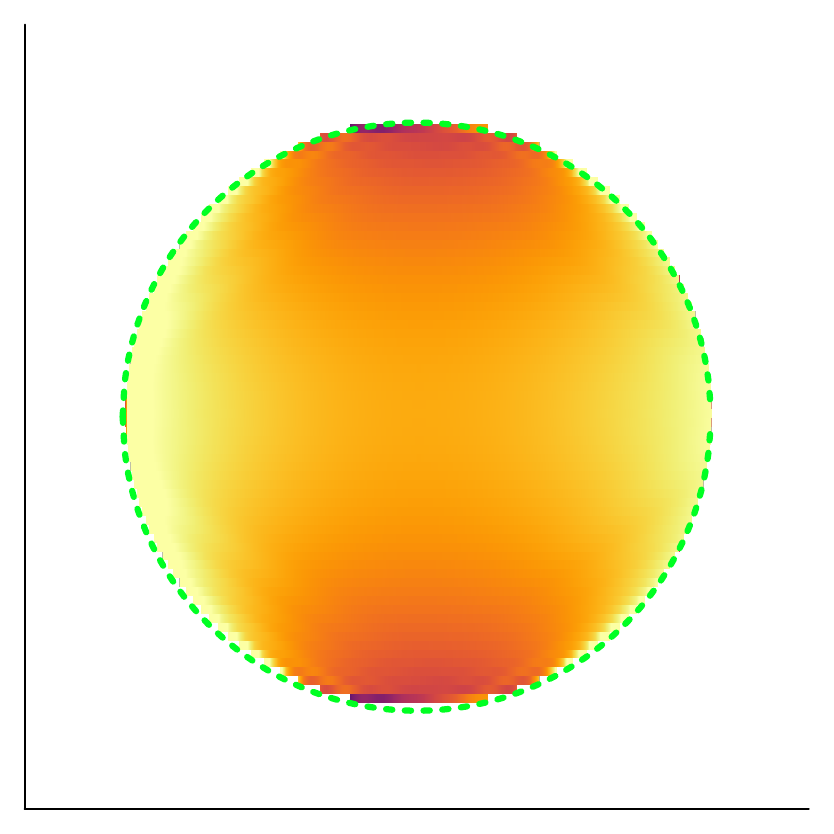}};
		\node at (-6.8,3) {$(a)$};
		\node at (0,3) {$(b)$};
		\node at (-4.2,-2.4) {$\tilde{x}$};
		\node at (2.6,-2.4) {$\tilde{x}$};
		\node at (-6.8,0.25) {$\tilde{y}$};
		\node at (0,0.25) {$\tilde{y}$};
		\node[rotate=90] at (-1,0.3) {$\mathrm{Re}(\hat{h})$};
		\node[rotate=90] at (5.6,1.45) {$\mathrm{Re}(\hat{Q})$};
		\node at (5.2,0.7) {$-0.5$};
		\node at (5.2,2.1) {$0.5$};
	\end{tikzpicture}
	\caption{(a) A plan view of the real part of the wave field produced by a circular source of radius $\pi/2$ at $v=1.25$. The blue dashed lines demonstrate that the wedge evolves along the characteristics \eqref{eq:sup_characteristics}. The inset shows the real part of the optimal source that occupies the circular region traced with the green circle on the main graph. (b) The same plot for $v=1.75$ illustrating that the wedge varies with velocity.}
	\label{fig:sup_graphs}
\end{figure}
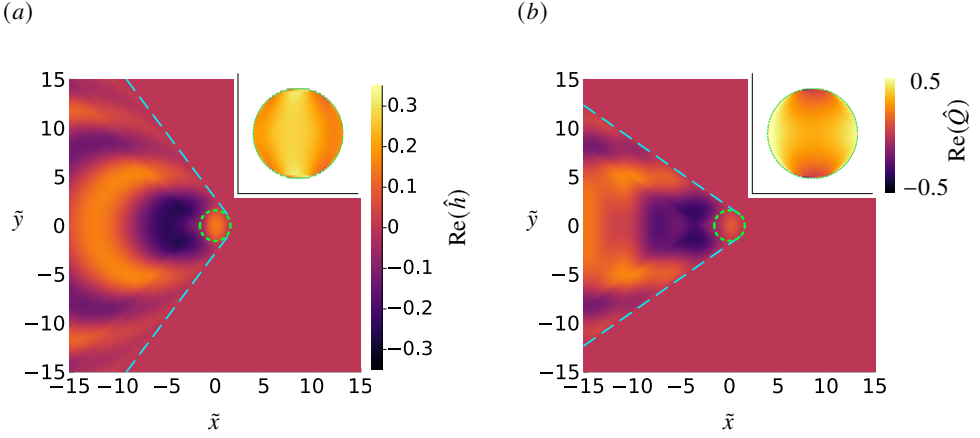

In figure \ref{fig:sup_graphs}, we see that the wedge behind the body does indeed expand parallel to \eqref{eq:sup_characteristics}. The difference in slope between figure \ref{fig:sup_graphs}$a$ and figure \ref{fig:sup_graphs}$b$ demonstrates the velocity dependence of the wedge angle. As expected, these solutions have a symmetry in $\tilde{y}$ to minimise thrust in the associated direction. 
Similar to the subcritical case, we can study the validity of the solution, using $k_{\mathrm{prox}}=\max|\widetilde{\nabla}\hat{h}|/\max|\hat{h}|$ as a measure of wave steepness. For a velocity of $1.05\leq v\leq2$ we find $k_{\mathrm{prox}}<1.7$ suggesting that the linear assumption remains valid for all supercritical velocities tested.

Inspired by the optimal solutions seen in the start-up and subcritical cases, we can check whether $\mathcal{L}_{2D}\hat{Q}=0$ holds in the supercritical regime with a circle of $r=\pi/2$.
Evaluated $3dy$ grid spaces radially inwards from the boundary of $\hat{Q}$, the residual is of order $10^{-2}$. However, the edge points do not satisfy the $\mathcal{L}_{2D}\hat{Q}=0$ very accurately (residual $\sim10^{1}$ to $10^{2}$). As discussed later, this may be due to a numerical instability at the edge of the source, but with a more powerful norm constraint achieved with $\beta=1,r=3\pi/4$ the error is of order $10^{-3}$ even at the boundaries.
In any case, we will attempt the analytically informed optimisation with separation of variables like the previous sections. 
Using $\hat{Q}$ from \eqref{eq:subandsup_sep_variables_result}, we assess the relative error between the two approaches to understand whether a source satisfying $\mathcal{L}_{2D}\hat{Q}=0$ can produce the same maximal thrust. For $r=\pi/2$, we find the relative error $\lesssim 15\%$ for thrust with velocities in the range $1.1\leq v\leq 1.45$ and increases to $\approx 50\%$ by $v=2$. However, for a larger body $(r=3\pi/4)$ the error reduces since the norm constraint is harsher leading to a relative error $\lesssim10\%$ for $1.1\leq v\leq2$.
The result does not support $\mathcal{L}_{2D}\hat{Q}=0$ to a high level of accuracy, but this could be due to the source being unstable at the edge. Greater violation of the condition $\mathcal{L}_{2D}\hat{Q}=0$, particularly at the sides and rear, is observed. Here, there is a spike in magnitude which aims to increase the amplitude of the resulting waves at the limits of the body. 
Perhaps a more accurate model applying a more restrictive norm constraint is required to investigate this further, but we will proceed with the present numerical scheme since this is not our focus.

\begin{figure}
	\centering
	\begin{tikzpicture}
		\node at (0,0) {\includegraphics[width=0.7\textwidth]{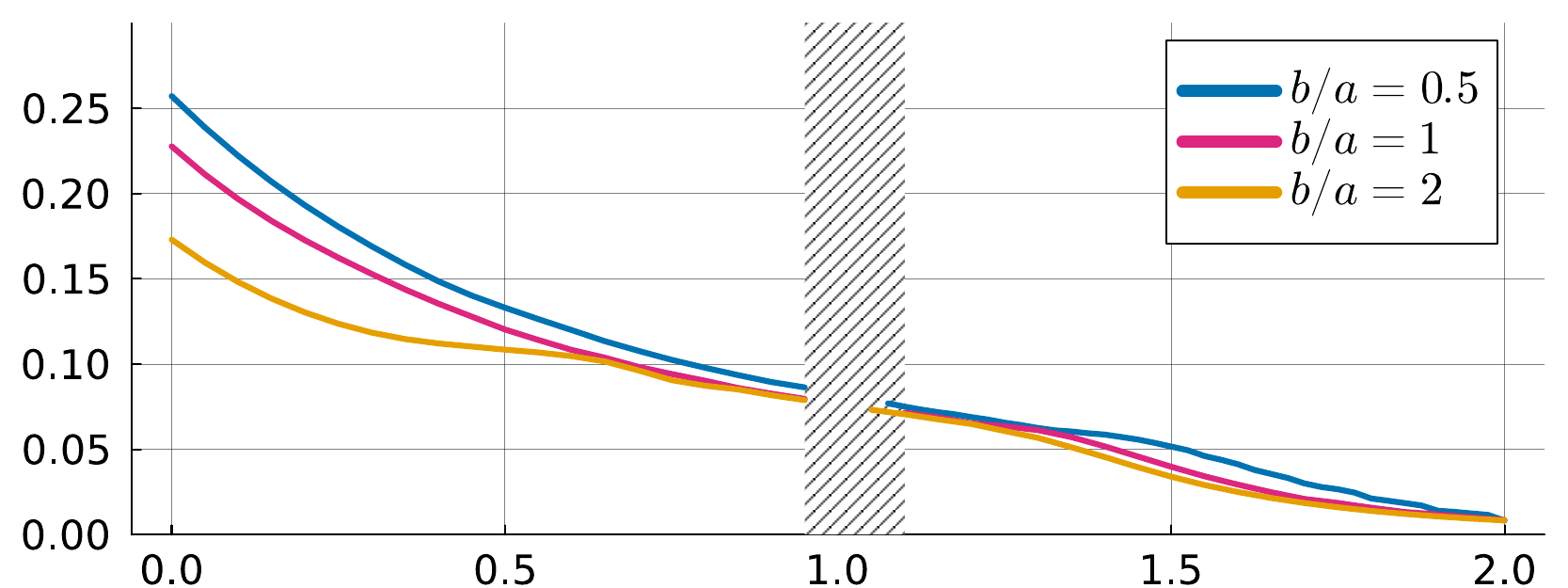}};	
		\node at (0.4,-2.1) {velocity ($v$)};
		\node[rotate=90] at (-5,0) {$\bar{F}_{T}$};
		\node at (0,-4.1) {\includegraphics[width=0.7\textwidth]{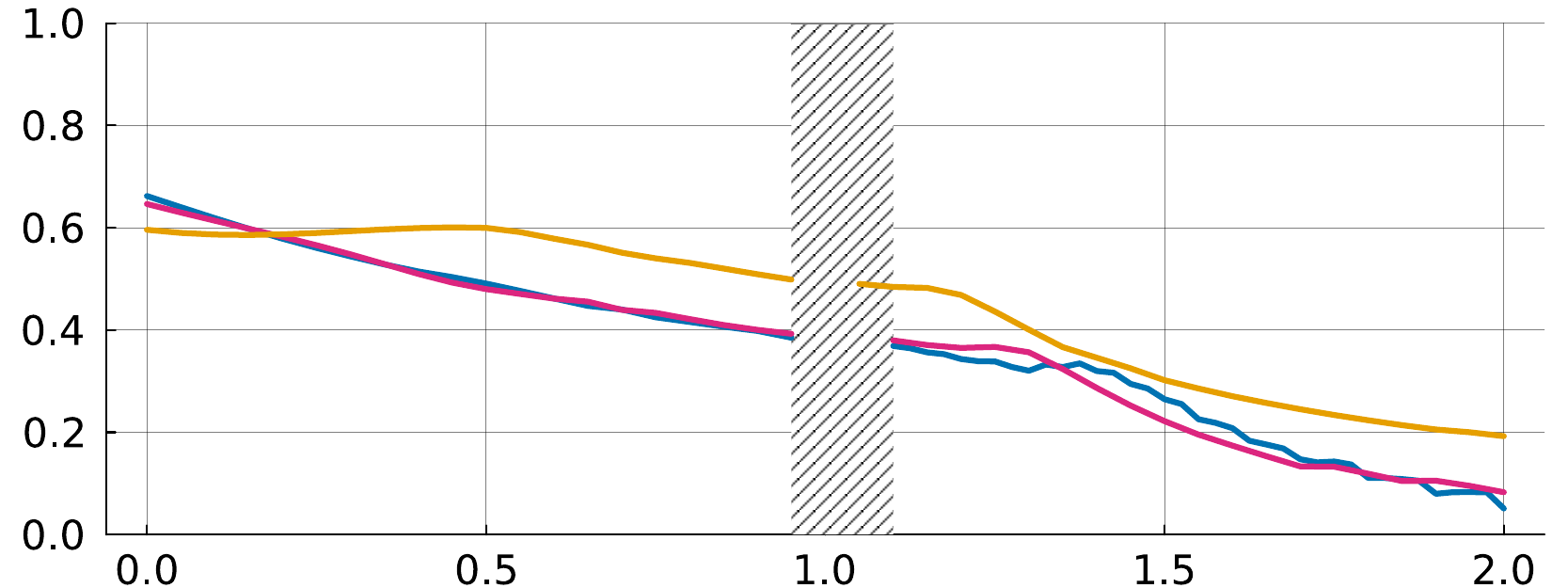}};	
		\node at (0.4,-2.3-4.1) {velocity ($v$)};
		\node[rotate=90] at (-5,-4.1) {$\eta$};
		\node at (-5,1.75) {$(a)$};
		\node at (-5,1.75-4.1) {$(b)$};
	\end{tikzpicture}
	\caption{$(a)$ Demonstration of the evolution of optimal thrust over subcritical and supercritical velocities for ellipses with three different aspect ratios $b/a$ and equal area equivalent to a circle of radius $\pi/2$. 
	$(b)$ A similar plot for the efficiency. The hatched region represents where the numerical scheme becomes inaccurate. The narrow body efficiency plot appears jagged due to computational limits on the grid. A grid of width $350$ was used for low supercritical velocities, but an increase in grid size is expected to smooth out the graph, as was the case for $b/a=1$.}
	\label{fig:full_v}
\end{figure}
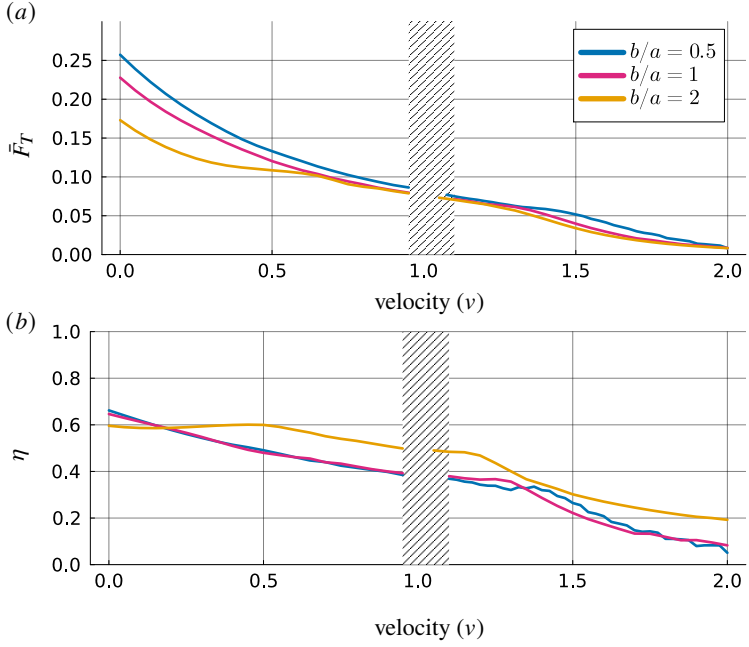

Like Section \ref{subsec:sub_aspectratio}, we evaluate maximal thrust over supercritical velocities for different aspect ratios ($b/a$) of an elliptical source. A similar preference towards a longer body ($a>b$) is found in the supercritical case. This is again interpreted as the source aiming to accommodate more wavelengths in the direction of travel. Rather than showing similar contour plots to figure \ref{fig:sub_sweeps}, we show the evolution of the maximal thrust with velocity in figure \ref{fig:full_v}$a$. As velocity increases thrust decreases, which may also support the argument that the amount of a wavelength the body can accommodate is crucial to thrust (since the Doppler shift stretches the wavelength). Interestingly, the wider body appears to be more efficient (figure \ref{fig:full_v}$b$). This makes intuitive sense since the body has a shorter side length available to produce efficiency reducing side waves. However, we note that the bounded norm case does not necessarily find maximum efficiency in WDP.

Given the intuition gained on the effect of aspect ratio, we will next take an exploratory look at general shape optimisation, where the source shape becomes a control.

\section{Optimising the shape of the source}\label{sec:optimal_shaping}

Previously, the source was prescribed as a fixed shape, such as a rectangle or an ellipse. We will now take an exploratory look at including the shape as an optimisation variable. We consider a circular region $\mathcal{D}$ of radius $r$ and constrain the source to be a fraction $n<1$ of the area of $\mathcal{D}$. This allowed area is calculated using a shape function that finds the points where $\hat{Q}\neq0$. 
Numerically, we smooth the shape function with a Gaussian distribution, and integrate over the region $\mathcal{D}$:
\begin{equation}
	\mathrm{Area} = \iint_{\mathcal{D}}\bigg( 1-\mathrm{e}^{-|\hat{Q}|^2/\sigma^2}\bigg)\,\mathrm{d}A,
\end{equation}
where $\sigma$ is a regularisation parameter analogous to the standard deviation.
Hence, the area constraint is  
\begin{equation}
	\mathrm{Area}\leq n\pi r^2,
\end{equation}
which permits the source to spread itself freely within the prescribed circular region. Meanwhile, the rest of the set-up remains the same ($\beta=1$), but we will discuss the subcritical and supercritical cases separately, displaying examples in both.
We will demonstrate the possibility for improvements by generalising the examples shown in Sections \ref{sec3:startup} and \ref{sec4:subcrit} (figures \ref{fig4:subcritical_numerical_result} and \ref{fig:sup_graphs}). 

\begin{figure}
	\centering
	\begin{tikzpicture}
		\node at (-3.8,0) {\includegraphics[width=0.4\textwidth]{fig8_cbar.pdf}};
		\node at (3.2,1.3) {\includegraphics[width=0.35\textwidth]{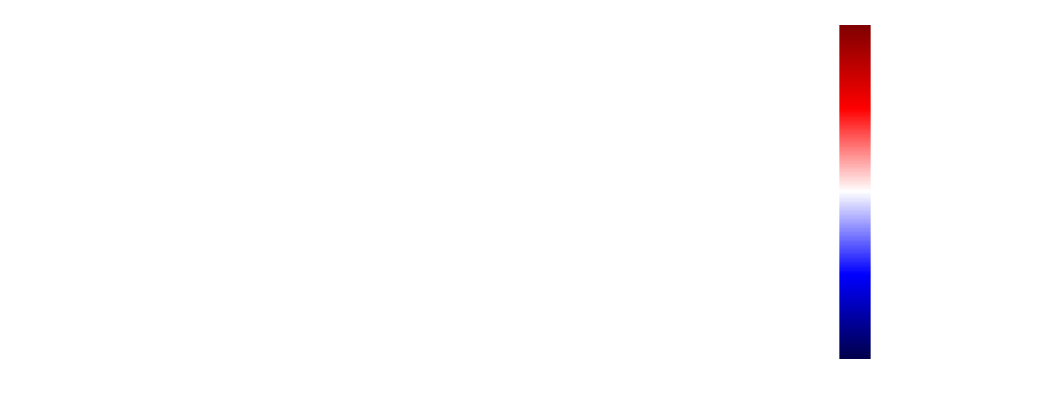}};
		\node at (-4.5,0) {\includegraphics[width=0.35\textwidth]{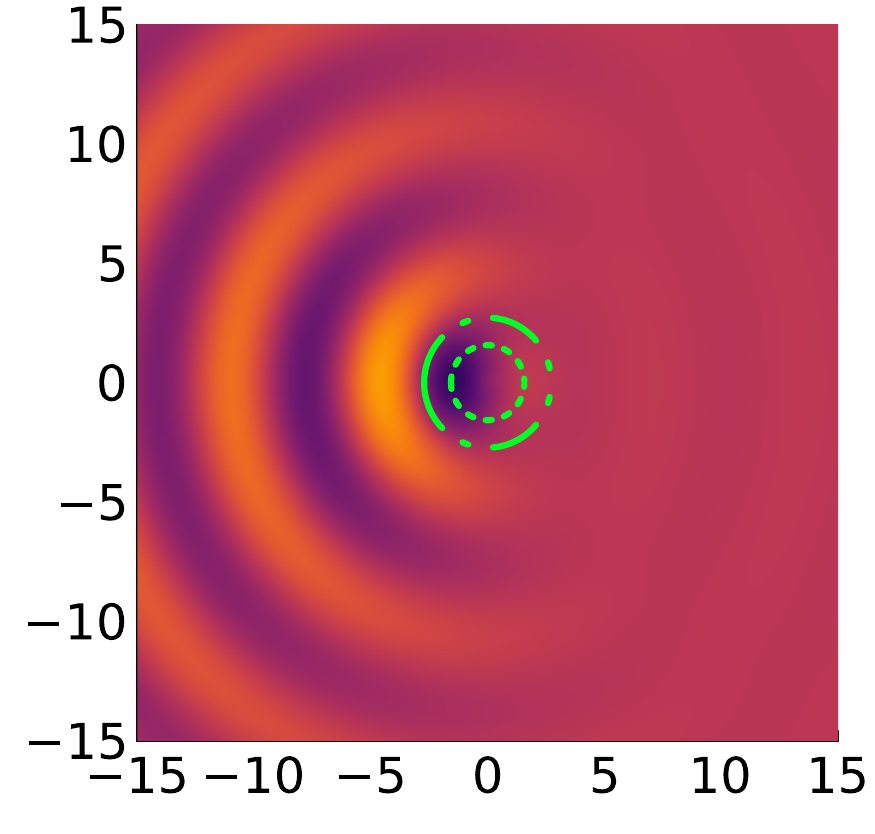}};
		\node at (2.3,0) {\includegraphics[width=0.35\textwidth]{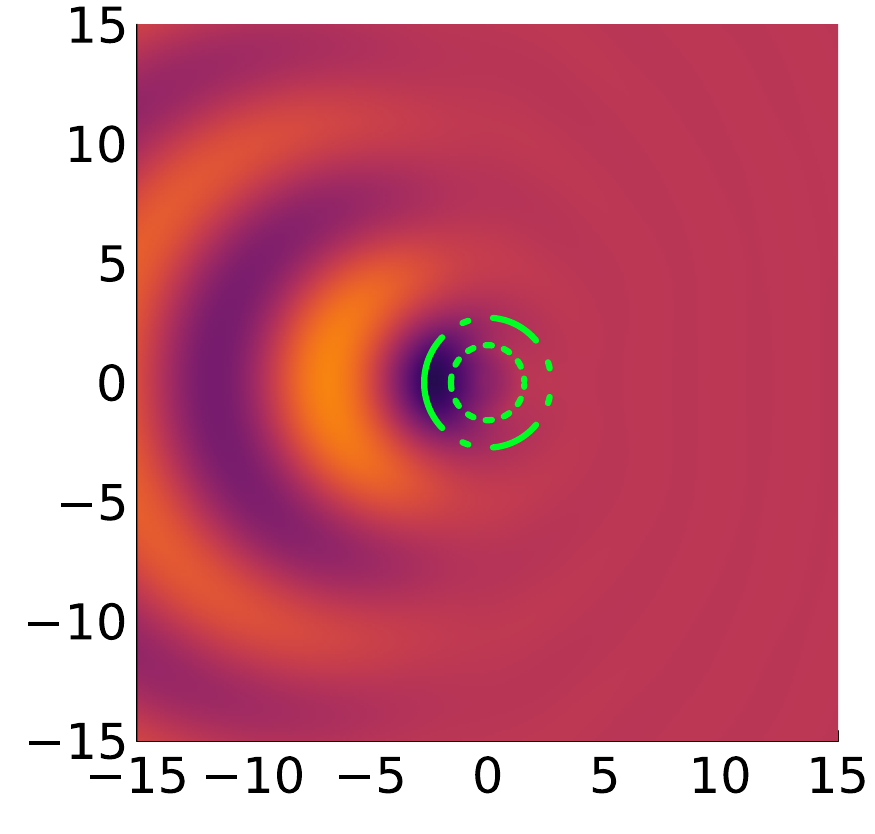}};
		\node at (-3,1.42) {\includegraphics[width=0.12\textwidth]{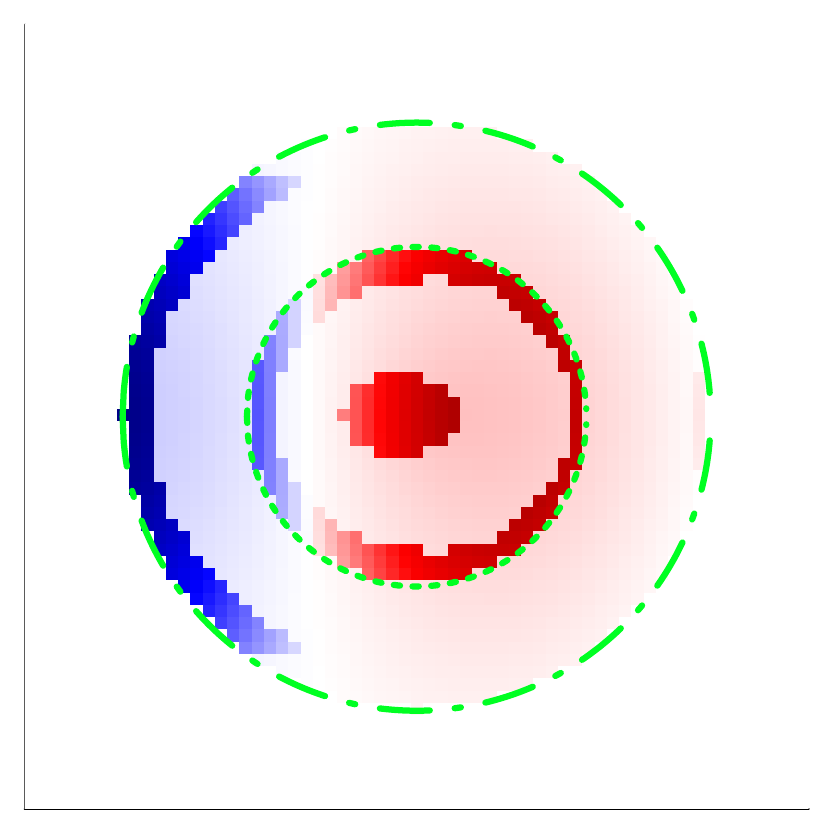}};
		\node at (3.8,1.42) {\includegraphics[width=0.12\textwidth]{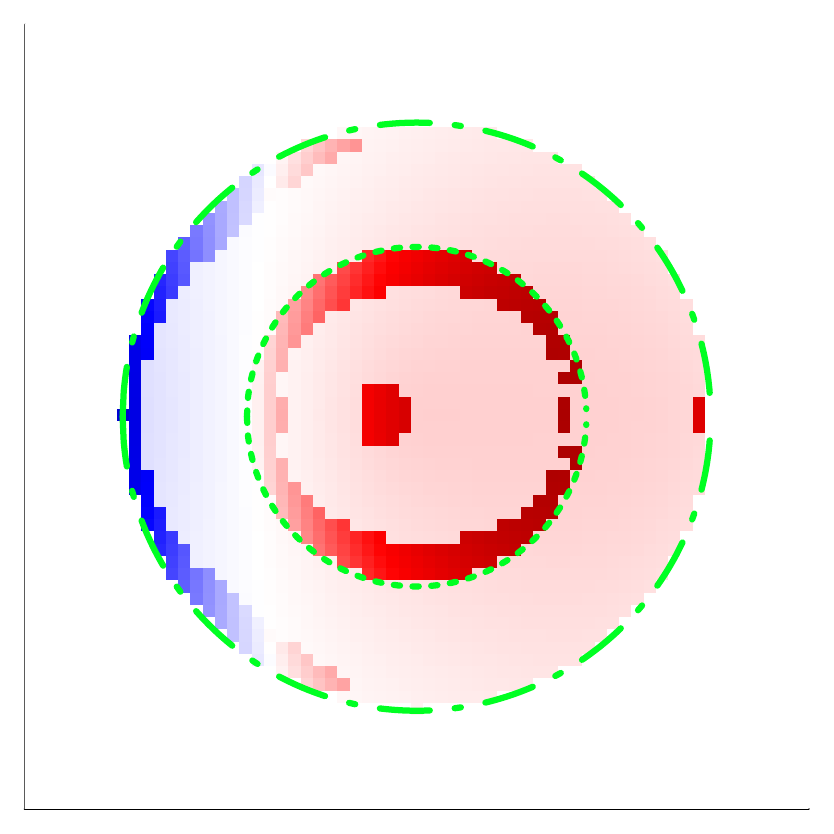}};
		\node at (-6.8,2.5) {$(a)$};
		\node at (0,2.5) {$(b)$};
		\node at (-4.2,-2.4) {$\tilde{x}$};
		\node at (2.6,-2.4) {$\tilde{x}$};
		\node at (-6.8,0.25) {$\tilde{y}$};
		\node at (0,0.25) {$\tilde{y}$};
		\node[rotate=90] at (-1,0.3) {$\mathrm{Re}(\hat{h})$};
		\node[rotate=90] at (5.6,1.45) {$\mathrm{Re}(\hat{Q})$};
		\node at (5.2,0.7) {$-0.6$};
		\node at (5.2,2.1) {$0.6$};
	\end{tikzpicture}
	\caption{(a) The resulting wave field produced by an optimal shape found when allowed to occupy $1/3$ the area of a circle of radius $r=\sqrt{3}\pi/2$ (top corner) at $v=0$ on a $240\times240$ grid. The $r=\sqrt{3}\pi/2$ region is represented by the outer dashed green circle while the circle of equal area $r=\pi/2$ is the inner dotted circle. (b) The result under the same conditions at $v=0.5$, as in figure \ref{fig4:subcritical_numerical_result}}
	\label{fig:v0_gen_shape}
\end{figure}

Beginning with the start-up case ($v=0$), we will allow our shape to occupy a third ($n=1/3$) of a circle of radius $r=\sqrt{3}\pi/2$ so that the resulting source has equal area to the $r=\pi/2$ circles in figures \ref{fig4:subcritical_numerical_result} and \ref{fig:sup_graphs}. The regularisation parameter $\sigma$ will be scaled with the uniform mesh ($dx$ or $dy$) to ensure the function is sufficiently smoothed.
The resulting optimal source is displayed in figure \ref{fig:v0_gen_shape}$a$. It appears that the source takes a hollow circular shape within the region which is partially filled by another circle. The remaining area allocation is spent on the edges. 
This is difficult to interpret, so we will limit our analysis to demonstrating the potential thrust and efficiency gains. 

The original circle produces a thrust of $\bar{F}_{T} \approx 0.2276$ with an efficiency of $\eta \approx 0.6465$. Meanwhile, the shaping result produces $\bar{F}_{T} \approx 0.2297$ with an efficiency of $\eta \approx 0.8045$. The optimal shaping produces a similar thrust with increased efficiency. We see similar behaviour when $v=0.5$ (figure \ref{fig:v0_gen_shape}$b$). A circle of the same area produces $\bar{F}_{T}\approx 0.1204$ with $\eta \approx 0.4808$ while the optimal shaping result produces $\bar{F}_{T}\approx 0.1283$ with $\eta\approx0.7212$. Again, we see a minor thrust improvement supplemented by a large efficiency gain. Therefore, expanding the influence of the source in patches over a larger area is an avenue to improve WDP (not dissimilar to a catamaran).

\begin{figure}
	\centering
	\begin{tikzpicture}
		\node at (-3.8,0) {\includegraphics[width=0.4\textwidth]{sup_fig_hbar.pdf}};
		\node at (3.2,1.3) {\includegraphics[width=0.35\textwidth]{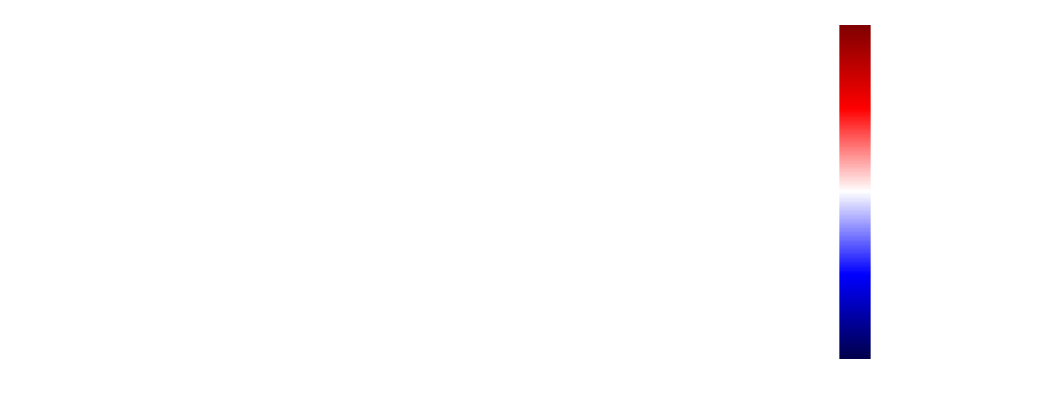}};
		\node at (-4.5,0) {\includegraphics[width=0.35\textwidth]{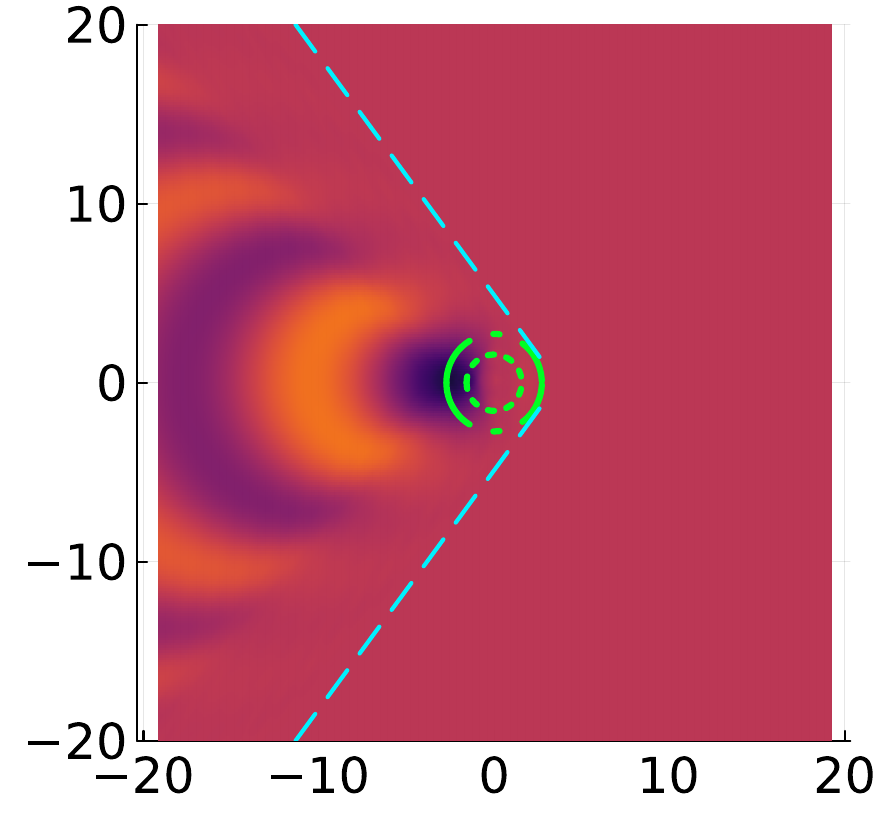}};
		\node at (2.3,0) {\includegraphics[width=0.35\textwidth]{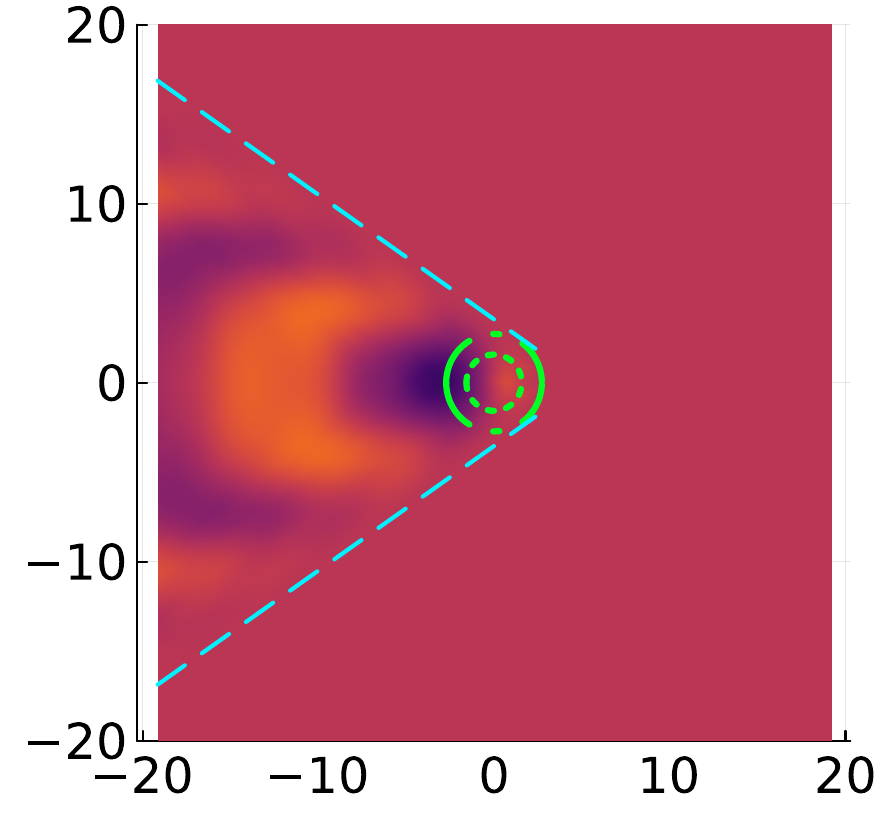}};
		\node at (-3.2,1.4) {\includegraphics[width=0.125\textwidth]{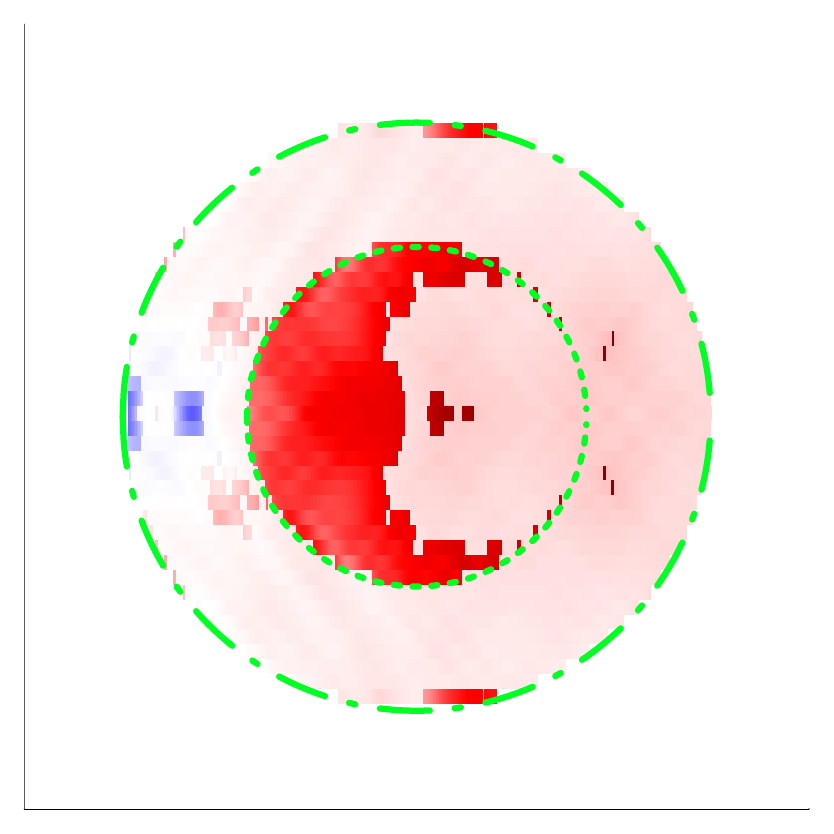}};
		\node at (3.6,1.4) {\includegraphics[width=0.125\textwidth]{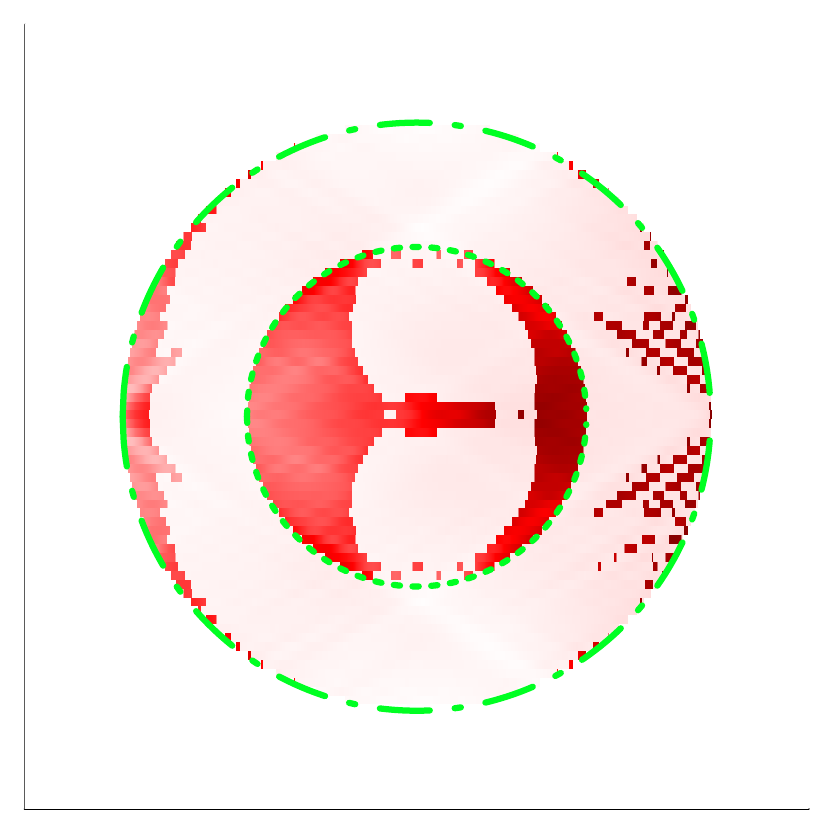}};
		\node at (-6.8,3) {$(a)$};
		\node at (0,3) {$(b)$};
		\node at (-4.2,-2.4) {$\tilde{x}$};
		\node at (2.6,-2.4) {$\tilde{x}$};
		\node at (-6.8,0.25) {$\tilde{y}$};
		\node at (0,0.25) {$\tilde{y}$};
		\node[rotate=90] at (-1,0.3) {$\mathrm{Re}(\hat{h})$};
		\node[rotate=90] at (5.6,1.45) {$\mathrm{Re}(\hat{Q})$};
		\node at (5.2,0.7) {$-0.6$};
		\node at (5.2,2.1) {$0.6$};
	\end{tikzpicture}
	\caption{(a) The wave field produced by the source that results from optimal shaping (equivalent to a circle $r=\pi/2$) (inset) for $v=1.25$ and a grid $250$ divisions wide (c.f. figure \ref{fig:sup_graphs}$a$). Note the differing colourbars. (b) A similar wave field for $v=1.75$ produced by the source seen in the top corner (c.f. figure \ref{fig:sup_graphs}$b$). Again the blue and green lines represent the characteristic trajectory \eqref{eq:sup_characteristics} and the position of the source, respectively.}
	\label{fig:sup_genshape}
\end{figure}

We can apply a similar approach to the supercritical case, where the velocities $v=1.25$ and $v=1.75$ will be discussed in comparison to Section \ref{sec6:supercrit}. The distance between discrete points in the $\tilde{x}$- and $\tilde{y}$-directions is not equal. Therefore, we scale $\sigma$ with the $\tilde{y}$-direction ($dy$), the larger of these distances. The optimal source occupying a third of the prescribed area ($n=1/3$) when $v=1.25$ is displayed with its wave field in figure \ref{fig:sup_genshape}$a$ while the case for $v=1.75$ is found in figure \ref{fig:sup_genshape}$b$. 
The source has a core circular shape, but forms quite a complicated shape elsewhere. 

We again compare the new shape's thrust and efficiency output to a circle of identical area. For $v=1.25$, the circle obtains $\bar{F}_{T}\approx0.0631$ with $\eta\approx0.3661$ while the shape shown achieves $\bar{F}_{T}\approx0.0752$ with $\eta\approx0.5052$. This is only a minor increase in thrust, but efficiency gains are still made. 
However, when we consider the same problem with $v=1.75$, the shape optimisation appears to improve both thrust and efficiency. The circle of equal area achieves $\bar{F}_{T}\approx0.0181$ with $\eta\approx0.1276$, while the new shape achieves $\bar{F}_{T}\approx0.0468$ with $\eta\approx0.3957$. The small thrust values are due to the Doppler shift at these speeds stretching the wavelength much larger than a body length. In this regime where the body cannot accommodate a significant proportion of a wavelength, optimal shaping appears to be an avenue for substantial gains.

Further efficiency gains may be found through a more refined optimal shaping approach, but the aim of this section is to demonstrate the possibilities that manipulating the shape can provide. One potential avenue to explore could be to look for harmonics like those seen in \citet{odonovan2026}. Each of the sources displayed has a main body part that has an approximate radius of $\pi/2$, the radius of the circle of equal area that forms the optimisation starting guess. The optimisation starts to then move some of this area elsewhere to optimise further. 
This may suggest a possible harmonic appearing, or else simply an artifact of multiple local maxima that are selected depending on choice of starting guess.
In figure \ref{fig:sup_genshape}$a$ the edge of the wedge has waves that are hard to discern on the heatmap. This region is composed of small amplitude side waves, suggesting another region that optimal shaping aims to reduce losses in efficiency. 

To supplement these results, we will conclude this section by discussing the limitations of our shaping approach. 
Firstly, in the subcritical case, we saw that the source has a tendency to trace the domain that it is permitted to occupy. 
Although this reaffirms the notion of increasing the number of wavelengths accommodated within the body,
a source entirely inside the domain could be explored by further restricting $\hat{Q}$. 
Secondly, the optimal solutions appeared to vary with the grid size. This could be due to the complex valued solutions being unique up to an arbitrary argument or the numerical optimisation finding a different local maximum in the solution set or a combination of both. 
The parameter $\sigma$ used to smooth the shape function was fixed to depend on grid spacing. Further refinement of this parameter to improve the optimal shaping result remains unexplored. 
With more constraints applied to the model fitting with a specific use case, the solution set would be reduced, perhaps subduing this effect.
Since we are taking a general exploration into optimal shaping, adding more constraints and refining the numerical approach will remain beyond the scope of the present study.

\section{Discussion}\label{sec:discuss}

In this paper, we began by defining thrust due to wave-driven propulsion resulting from surface gravity waves. Applying a similar approach to \citet{odonovan2026}, we derive thrust (in the moving frame) decomposed into the direction of travel ($\tilde{x}$) and perpendicular to this ($\tilde{y}$), where the latter is shown to be zero when the wave field is symmetric about the $\tilde{x}$-axis. When travelling at subcritical velocities ($v<1$), we addressed the challenge of defining Sommerfeld boundary conditions for a source producing waves that are Doppler shifted in the moving frame, by exploiting the invariance of the wave equation under a Lorentz transformation. 
For supercritical velocities ($v>1$), the Doppler shift produces a wedge due to causality, and we demonstrate the velocity dependent slope of this wedge.
Using this set-up we pose an optimisation problem to understand how maximal thrust depends on the shape of the body. 
With a regularisation constraint on the norm of the control function and a fixed area, we find that a longer body has a larger maximal thrust for both subcritical and supercritical velocities while a wider body appears to promote efficiency. 
Following this, we demonstrate the possibility of optimising the shape of the body for further gains in thrust and efficiency by reducing wave momentum in the $\tilde{y}$-direction. 

These results are limited by the shallow water assumption, which is linearised to only study small amplitude oscillations. This allows us to focus on the contribution to WDP from surface waves. Therefore, the model does not account for any subsurface effects. Our choice to focus on a canoe sized raft allowed us to also neglect viscosity and surface tension. We would need to include such effects to optimise smaller scale bodies such as capillary surfers \citep{oza_theoretical_2023}. Similarly, our irrotational shallow water model cannot quantify the role of vortex-driven propulsion (VDP) relative to WDP as done in the case of the water strider \citep{Steinmann2021,GAO_FENG_2011,BUHLER_2007}. 

Within the limitations of the current set-up, we can pose further optimisation problems, primarily bounding the power \citep{odonovan2026}. This case would maximise efficiency by fixing the power injected. This was attempted in the current set-up, but leads to the source seeking to become highly discontinuous and the optimisation failing to converge. Perhaps this is in an effort to remove the waves in the $\tilde{y}$-direction entirely (not possible). A loose bounded norm constraint could be imposed to reduce this effect, but was not explored in this paper. To probe optimising efficiency, more complex numerical schemes could be applied. The numerical approach was kept simple to speed up the optimisation, but we could consider other approaches applied to the convected Helmholtz equation from aeroacoustics for the subcritical case \citep{marchner2021_PML_using_lorentz,hu2019_prandtl,barucq2022_boundary_conds}. 
The higher order absorbing boundary conditions described by \citet{barucq2022_boundary_conds} can also improve accuracy. Although the low-order and plane wave boundary conditions appear to differ from those in this paper, they apply to the same Green's function solution on a differently shaped domain with opposing sign conventions. 
Similarly, schemes that apply the $\tilde{y}$ boundary condition at the wedge, rather than the edge of the domain in the supercritical case, would prevent the possibility of numerical diffusion past the wedge that smoothes out any discontinuity here. Our approach was sufficient for the goal of studying shape, so this was not explored. 

This paper provides a template where the thrust and boundary conditions found can be used to guide similar approaches in more complex problems. The first step would be to remove the shallow water assumption and add depth to the problem. Retaining our focus on gravity waves, we can neglect viscosity and apply a potential flow approach, similar to the work of \citet{Benham2024propulsion}, and then expand this to three dimensions \citep{keeler2025_wake} to understand how subsurface flows affect our current results. Similar to \citet{Benham2024propulsion}, the flow can be coupled to the movement of a solid body at the surface. It would be interesting to compare the thrust and efficiency gains from optimal shaping in this case and explore shapes and layouts of raft sections \citep{boucher2018_thin,Tuck1998OptimumHS}. Increasing the complexity further to more realistic flows with CFD simulations would offer the platform to study WDP along with VDP similar to \citet{Steinmann2021}. However, the number of constraints increases with these set-ups, so the work in this paper will be vital in selecting variables to optimise, where effects like dispersive waves and dissipation complicate the results. 

The wedge shaped supercritical wake pattern may differ to a conventional body on the fluid surface, but it does provoke a future direction to study another area beyond the scope of this work.
The thrust produced will balance drag which includes wave drag from the surface waves produced by the body \citep{boucher2018_thin,Tuck1998OptimumHS,Dode2022,keeler2025_wake,Yuan2021ducks}. With the addition of drag to a WDP optimisation problem, there would be an interesting interplay between minimising wave drag \citep{boucher2018_thin} and maximising WDP, both of which are dependent on the waves produced.
Along with leading to these future directions, the work presented in this paper further builds our understanding of the mechanisms involved in WDP across the fluid surface to maximise its role in propulsion.  

\vspace{1em}
\noindent\textbf{Funding} GP Benham is funded by the European Union (ERC, SurFSUP, 101219032). Views and opinions expressed are however those of the author(s) only and do not necessarily reflect those of the European Union or the European Research Council. Neither the European Union nor the granting authority can be held responsible for them. 

\vspace{1em}
\noindent\textbf{Declaration of interests.} The authors report no conflict of interest.

\vspace{1em}
\noindent\textbf{Data availability statement.} Examples of the code used in this paper are found at \url{https://github.com/daireodonovan/shape-in-wdp}.

\appendix

\section{Thrust derivation}\label{ap_sec:thrust}

We begin with the dimensionless wave equation in the stationary frame 
\begin{equation}
	\nabla^2 h - \frac{\partial^2h}{\partial t^2} = -Q(x,y,t).
\end{equation}
The Galilean transformation \eqref{eq:galilean_transformation} is taken to operate in the moving frame, where the $\tilde{x}$-axis is fixed in the direction of travel,
\begin{equation}\label{ap_eq:wave_eq_moving}
	(1-v^2)\frac{\partial^2h}{\partial \tilde{x}^2} + 2v\frac{\partial^2h}{\partial \tilde{x}\,\partial\tilde{t}} + \frac{\partial^2 h}{\partial \tilde{y}^2} -\frac{\partial^2 h}{\partial \tilde{t}^2} = -Q.
\end{equation}
We propose that thrust per unit length is written as a vector decomposed into $\tilde{x},\tilde{y}$ components. 
\begin{equation}\label{ap_eq:thrust_pde_system}
	\left((1-v^2)\frac{\partial^2h}{\partial \tilde{x}^2} + 2v\frac{\partial^2h}{\partial \tilde{x}\,\partial\tilde{t}} + \frac{\partial^2 h}{\partial \tilde{y}^2} -\frac{\partial^2 h}{\partial \tilde{t}^2}\right)\begin{pmatrix} h_{\tilde{x}} \\ h_{\tilde{y}}\end{pmatrix} = -Q\begin{pmatrix} h_{\tilde{x}} \\ h_{\tilde{y}}\end{pmatrix}.
\end{equation}
This system of PDEs defines thrust in the $\tilde{x}$- and $\tilde{y}$-directions, respectively.

\subsection{Thrust in the x-direction}\label{ap_sec:xthrust}

Taking a surface integral of the first element of \eqref{ap_eq:thrust_pde_system},
\begin{equation}
	\begin{aligned}
	\int^{\tilde{x}^{+}}_{\tilde{x}^{-}}\int^{\tilde{y}^{+}}_{\tilde{y}^{-}} \left((1-v^2)h_{\tilde{x}\tilde{x}}h_{\tilde{x}} +2vh_{\tilde{x}\tilde{t}}h_{\tilde{x}} + h_{\tilde{y}\tilde{y}}h_{\tilde{x}} - h_{\tilde{t}\tilde{t}}h_{\tilde{x}}\right)\,\mathrm{d}\tilde{x}\,\mathrm{d}\tilde{y} 
	\\
	= -\int^{\tilde{x}^{+}}_{\tilde{x}^{-}}\int^{\tilde{y}^{+}}_{\tilde{y}^{-}} Q\,h_{\tilde{x}}\,\mathrm{d}\tilde{x}\,\mathrm{d}\tilde{y},
	\end{aligned}
\end{equation}
where $\tilde{x}^{\pm},\tilde{y}^{\pm}$ represent the edges of a square region traced in the far-field away from the body, we assume the body oscillates periodically like \eqref{eq:periodic_assumption} and take the time-average over a period of oscillation,
\begin{equation}\label{ap_eq:time_avg_xthrust_start}
\begin{gathered}
	\frac{1}{4}\int^{\tilde{x}^{+}}_{\tilde{x}^{-}}\int^{\tilde{y}^{+}}_{\tilde{y}^{-}}\bigg((1-v^2)\left(\hat{h}^{*}_{\tilde{x}\tilde{x}}\hat{h}_{\tilde{x}}+\hat{h}_{\tilde{x}\tilde{x}}\hat{h}_{\tilde{x}}^{*}\right) 
	+ \left(\hat{h}_{\tilde{y}\tilde{y}}^{*}\hat{h}_{\tilde{x}}+\hat{h}_{\tilde{y}\tilde{y}}\hat{h}_{\tilde{x}}^{*}\right) 
	+ \left(\hat{h}^{*}\hat{h}_{\tilde{x}}+\hat{h}\hat{h}_{\tilde{x}}^{*}\right)\bigg)\,\mathrm{d}\tilde{x}\,\mathrm{d}\tilde{y} 
	\\
	= -\frac{1}{4}\int^{\tilde{x}^{+}}_{\tilde{x}^{-}}\int^{\tilde{y}^{+}}_{\tilde{y}^{-}} \hat{Q}^{*}\hat{h}_{\tilde{x}}+\hat{Q}\hat{h}_{\tilde{x}}^{*}\,\mathrm{d}\tilde{x}\,\mathrm{d}\tilde{y} = -\langle\hat{Q},\hat{h}_{\tilde{x}}\rangle.
\end{gathered}
\end{equation}
Note that the cross term ($2vh_{\tilde{x}\tilde{t}}$) cancels out when the time-average is taken. To simplify, some terms can be integrated,
\begin{equation}
	-\langle\hat{Q},\hat{h}_{\tilde{x}}\rangle
	=
	\frac{1}{4}\int^{\tilde{y}^{+}}_{\tilde{y}^{-}}\left[(1-v^2)|\hat{h}_{\tilde{x}}|^2 + |\hat{h}|^2\right]_{\tilde{x}^{-}}^{\tilde{x}^{+}}\,\mathrm{d}y + \frac{1}{4}\int^{\tilde{x}^{+}}_{\tilde{x}^{-}}\int^{\tilde{y}^{+}}_{\tilde{y}^{-}}\left(\hat{h}_{\tilde{y}\tilde{y}}^{*}\hat{h}_{\tilde{x}}+\hat{h}_{\tilde{y}\tilde{y}}\hat{h}_{\tilde{x}}^{*}\right)\,\mathrm{d}\tilde{x}\,\mathrm{d}\tilde{y}. 
	\label{ap_eq:xthrust_equality_before_parts}
\end{equation}
To maintain a general thrust equality, we will not use any boundary conditions to simplify the terms on the right-hand side. These terms are analogous to what were used by \citet{odonovan2026}, and quantify the flux of waves out in the $\tilde{x}$-direction contributing to the thrust. Integration by parts is used to write the final integral with boundary terms:
\begin{equation}\label{ap_eq:thrust_gen_expression}
	\langle\hat{Q},\hat{h}_{\tilde{x}}\rangle
	=
	-\frac{1}{4}\int^{\tilde{y}^{+}}_{\tilde{y}^{-}}\left[(1-v^2)|\hat{h}_{\tilde{x}}|^2
	+ |\hat{h}|^2 
	- |\hat{h}_{\tilde{y}}|^2\right]_{\tilde{x}^{-}}^{\tilde{x}^{+}}\,\mathrm{d}\tilde{y} 
	- \frac{1}{4}\int^{\tilde{x}^{+}}_{\tilde{x}^{-}}\bigg[\hat{h}_{\tilde{y}}^{*}\hat{h}_{\tilde{x}}+\hat{h}_{\tilde{y}}\hat{h}_{\tilde{x}}^{*}\bigg]^{\tilde{y}^{+}}_{\tilde{y}^{-}}\,\mathrm{d}\tilde{x}. 
\end{equation}
This is interpreted as an equality between thrust due to momentum injected by $\hat{Q}$ and that which is radiated away by the wave field. Therefore, we claim the time-averaged thrust is:
\begin{equation}\label{ap_eq:x_thurst_defn}
	\bar{F}_{T} = \langle \hat{Q},\hat{h}_{\tilde{x}}\rangle. 
\end{equation}
Each term on the right-hand side is evaluated at either the $\tilde{x}$ or $\tilde{y}$ boundary. This is already looking indicative of the radiation stress mechanisms discussed by \citet{Longuet-Higgins1977meanforces}. 
The thrust appears to be modulated by the movement of the wave in the $\tilde{x}$-,$\tilde{y}$-direction with the $\hat{h}_{\tilde{x}}$ and $\hat{h}_{\tilde{y}}$ flux terms. To understand the terms in \eqref{ap_eq:thrust_gen_expression}, we will prescribe:
\begin{equation}
    h(\tilde{\boldsymbol{x}},\tilde{t}) = \mathrm{Re}\left(\mathrm{e}^{i(\boldsymbol{k}\cdot\boldsymbol{\tilde{x}}-\tilde{t})}\right) \implies \hat{h} = \mathrm{e}^{i\boldsymbol{k}\cdot\boldsymbol{\tilde{x}}},
\end{equation}
a plane wave of unit amplitude in the direction of $\boldsymbol{k}=(k_{(\tilde{x})},k_{(\tilde{y})})$. A dispersion relation can be recovered by substituting into the homogeneous part of \eqref{ap_eq:wave_eq_moving},
\begin{equation}\label{ap_eq:pl_wave_dispersion}
	(1-v^2)k_{(\tilde{x})}^{2} - 2vk_{(\tilde{x})} + k_{(\tilde{y})}^{2} = 1,
\end{equation}
which we will study geometrically.

\begin{figure}
	\centering
	\begin{tikzpicture}[scale=0.8]
		\draw [bend right,red,thick] (1,0) to (0.9,0.55);
		\draw [bend right,red,thick] (10,0) to (8,0.4);
		\draw[ultra thick] (1-1,0) ellipse (2cm and 1cm);
		\draw[thick] (0-1,-3.5) -- (0-1,3.5);
		\draw[thick] (-2-1,0) -- (4-1,0);
		\draw[red, very thick] (-1,0) -- (2.42-1,0.7);
		\fill[red] (2.42-1,0.7) circle (2pt);
		\node[red, above right] at (0,-0.1) {$\alpha$};
		\node[right] at (0-1,3.5) {$k_{(\tilde{y})}$};
		\node[below] at (4-1,0) {$k_{(\tilde{x})}$};
		\node[red,above right] at (2.42-1, 0.7) {$\textbf{k}$};
		\draw[scale=0.6,domain=1.09:5.22,smooth,variable=\t,ultra thick] plot ({15+(1/(1-1.5*cos(\t r)))*cos(\t r)},{(1/(1-1.5*cos(\t r)))*sin(\t r)});
		\draw[scale=0.6,domain=5.55:7,smooth,variable=\t,ultra thick] plot ({15+(1/(1-1.5*cos(\t r)))*cos(\t r)},{(1/(1-1.5*cos(\t r)))*sin(\t r)});
		\draw[thick] (9,-3.5) -- (9,3.5);
		\draw[thick] (9-4,0) -- (9+2,0);
		\draw[red,very thick] (9+0,0) -- (9-3,1.2);
		\fill[red] (9-1.41,0.57) circle (2pt);
		\fill[red] (9-0.23,0.10) circle (2pt);
		\node[red, below left] at (9-1.31,0.7) {$\textbf{k}^{(2)}$};
		\node[red,below left] at (8.8,-0.1) {$\textbf{k}^{(1)}$};
		\node[red, above right] at (8.9,-0.1) {$\alpha$};
		\node[right] at (9+0,3.5) {$k_{(\tilde{y})}$};
		\node[below] at (9+2,0) {$k_{(\tilde{x})}$};
		\node at (-3,3.75) {$(a)$};
		\node at (4,3.75) {$(b)$};
	\end{tikzpicture}
	\caption{(a) Demonstration of a line of angle $\alpha$ starting at the focal point and intersecting with the subcritical ellipse of possible wavenumbers to give the plane wavenumber $\boldsymbol{k}$. (b) A similar demonstration for supercritical velocities where we now have a hyperbola which can have multiple intersections for certain angles $\alpha$.}
	\label{fig:wavenumber_demos}
\end{figure}
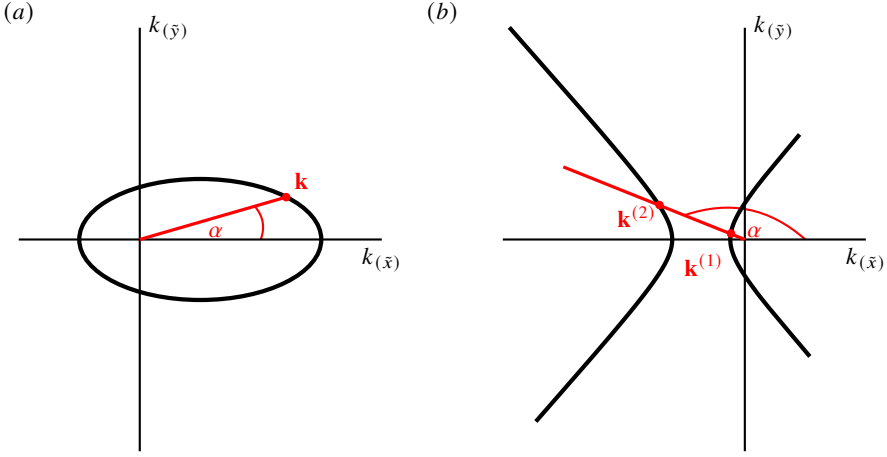

When $v<1$, \eqref{ap_eq:pl_wave_dispersion} is an ellipse,
\begin{equation}
	\frac{\left(k_{(\tilde{x})} - \frac{v}{1-v^2}\right)^2}{\left(\frac{1}{1-v^2}\right)^{2}} + \frac{k_{(\tilde{y})}^2}{\left(\frac{1}{1-v^2}\right)}=1.
\end{equation}
When $v=0$, we have a unit circle centred at the origin. However, the centre point moves from the origin when $v\neq0$, but one focal point remains at the origin. We will use this to set up a polar coordinate system to write the thrust in terms of angles. A point $\boldsymbol{k}$ on the ellipse is written in terms of the angle $\alpha$ it makes with the $k_{(\tilde{x})}$-axis:
\begin{equation}\label{eq:angle_decomp_coords_plane_waves}
	(k_{(\tilde{x})},k_{(\tilde{y})}) = \left(\frac{\cos\alpha}{1-v\cos\alpha},\frac{\sin\alpha}{1-v\cos\alpha}\right), 
\end{equation}
which is displayed in figure \ref{fig:wavenumber_demos}$a$. 
Applying the prescribed $\hat{h}$ to \eqref{ap_eq:thrust_gen_expression}, 
we can write: 
\begin{multline}\label{ap_eq:thrust_angle_form}
    \bar{F}_{T} = 
    -\frac{1}{4}\int^{\tilde{y}^{+}}_{\tilde{y}^{-}}\left[\left(1+\frac{(1-v^2)\cos^{2}\alpha}{(1-v\cos\alpha)^2}-\frac{\sin^2\alpha}{(1-v\cos\alpha)^2}\right)|\hat{h}|^2\right]_{\tilde{x}^{-}}^{\tilde{x}^{+}} \,\mathrm{d}\tilde{y} 
	\\
	- \frac{1}{2}\int^{\tilde{x}^{+}}_{\tilde{x}^{-}}\left[\left(\frac{\tan\alpha\cos^2\alpha}{1-v\cos\alpha}\right)|\hat{h}|^2\right]^{\tilde{y}^{+}}_{\tilde{y}^{-}}\,\mathrm{d}\tilde{x}.
\end{multline}
Firstly, if we take $\alpha=0$ at $\tilde{x}^{+}$ and $\alpha=\pi$ at $\tilde{x}^{-}$, we recover the thrust found in \citet{odonovan2026} integrated over $\tilde{y}$
\begin{equation}\label{ap_eq:one_d_thrust_analogue}
    \bar{F}_{T} = 
	-\frac{1}{2}\int^{\tilde{y}^{+}}_{\tilde{y}^{-}}\bigg[ \frac{1}{1-v}|\hat{h}_{+}|^2 - \frac{1}{1+v} |\hat{h}_{-}|^2 \bigg]\,\mathrm{d}\tilde{y}. 
\end{equation}
In general, we interpret \eqref{ap_eq:thrust_angle_form} as follows: in the presence of an asymmetry, the first integral has the difference in fore-aft wave amplitude squared. Since we are operating in two dimensions there is a correction dependent on the angle of the wave propagation which accounts for the fact that the $\tilde{y}$ component of the wave momentum cannot contribute to body momentum in the $\tilde{x}$-direction.
The second integral is to factor in the flux of momentum out of the sides of the source and since $k_{(\tilde{x})}$ is contributing to the thrust, its effect is being modulated by its ratio to $k_{(\tilde{y})}$ ($\tan\alpha$). 

When $v>1$, the dispersion relation results in a hyperbola (figure \ref{fig:wavenumber_demos}$b$),
Again, we use the focal point at $(0,0)$ to set up the coordinate system in terms of an angle using \eqref{eq:angle_decomp_coords_plane_waves}.
This will again result in \eqref{ap_eq:thrust_angle_form} with the same interpretation. 
However, there is a discontinuity in the hyperbola \eqref{eq:angle_decomp_coords_plane_waves} when $\alpha = \cos^{-1}(1/v),\in(0,\pi/2)$, so we split our angle into two separate regions. 
For $\alpha\in(\cos^{-1}(1/v),2\pi-\cos^{-1}(1/v))$ the front branch of the hyperbola is intersected once and additionally the rear branch is intersected for the subinterval $\alpha\in(\pi-\cos^{-1}(1/v),\pi+\cos^{-1}(1/v))$. Therefore, for $\alpha\in(-\cos^{-1}(1/v),\cos^{-1}(1/v))$ there are no intersections anywhere.
To represent this, we will rewrite \eqref{eq:angle_decomp_coords_plane_waves} in terms of the wavenumbers for each branch separately:
\begin{align}\label{ap_eq:sup_angle_1}
	\left(k^{(1)}_{\tilde{x}},k^{(1)}_{\tilde{y}}\right) = 
	\left(\frac{\cos\alpha}{1-v\cos(\alpha)}, \frac{\sin\alpha}{1-v\cos(\alpha)}\right), & & \alpha\in\left(\cos^{-1}\left(\frac{1}{v}\right),2\pi -\cos^{-1}\left(\frac{1}{v}\right)\right),
	\\
	\label{ap_eq:sup_angle_2}
	\left(k^{(2)}_{\tilde{x}},k^{(2)}_{\tilde{y}}\right) = 
	-\left(\frac{\cos\alpha}{1+v\cos(\alpha)}, \frac{\sin\alpha}{1+v\cos(\alpha)}\right), & & \alpha\in\left(\pi-\cos^{-1}\left(\frac{1}{v}\right),\pi+\cos^{-1}\left(\frac{1}{v}\right)\right).
\end{align}
If we take $\alpha=\pi$, we will get two wavenumbers which correspond to waves travelling with angle $0$ and $\pi$ in the rest frame. These are the fore--aft waves being Doppler shifted backwards from the one-dimensional case. Therefore, with the multivalued wavenumber expression, we again recover the one-dimensional thrust when $v>1$ and $\alpha=\pi$. 

\subsection{Thrust in the y-direction}\label{ap_sec:ythrust}

Applying the same methods used to derive the thrust in Appendix \ref{ap_sec:xthrust}, the time-averaged equality analogous to \eqref{ap_eq:time_avg_xthrust_start} is found in the $\tilde{y}$-direction.
Taking similar simplification steps produces:
\begin{equation}\label{ap_eq:ythrust_timeavg}
\begin{aligned}
	-\langle\hat{Q},\hat{h}_{\tilde{y}}\rangle=&
	\frac{1}{4}\int^{\tilde{x}^{+}}_{\tilde{x}^{-}}
	\left[|\hat{h}_{\tilde{y}}|^2 
	+ |\hat{h}|^2 
	- (1-v^2)|\hat{h}_{\tilde{x}}|^2\right]_{\tilde{y}^{-}}^{\tilde{y}^{+}}\,\mathrm{d}\tilde{x}
	\\ 
	&+ \frac{1-v^2}{4}\int^{\tilde{y}^{+}}_{\tilde{y}^{-}}\bigg[\hat{h}_{\tilde{y}}^{*}\hat{h}_{\tilde{x}}+\hat{h}_{\tilde{y}}\hat{h}_{\tilde{x}}^{*}\bigg]^{\tilde{x}^{+}}_{\tilde{x}^{-}}\,\mathrm{d}\tilde{y} 
	-\frac{iv}{2}\int^{\tilde{x}^{+}}_{\tilde{x}^{-}}\int^{\tilde{y}^{+}}_{\tilde{y}^{-}}\left(\hat{h}_{\tilde{x}}\hat{h}^{*}_{\tilde{y}}-\hat{h}_{\tilde{x}}^{*}\hat{h}_{\tilde{y}}\right)\,\mathrm{d}\tilde{x}\,\mathrm{d}\tilde{y}.
\end{aligned} 
\end{equation}
Unlike \eqref{ap_eq:thrust_gen_expression}, the cross term does not cancel. This is simplified using Green's theorem to evaluate at the boundaries,
\begin{equation}\label{ap_eq:ythrust_tidy}
\begin{aligned}
	-\langle\hat{Q},\hat{h}_{\tilde{y}}\rangle=&
	\frac{1}{4}\int^{\tilde{x}^{+}}_{\tilde{x}^{-}}
	\left[|\hat{h}_{\tilde{y}}|^2 
	+ |\hat{h}|^2 
	- (1-v^2)|\hat{h}_{\tilde{x}}|^2\right]_{\tilde{y}^{-}}^{\tilde{y}^{+}}\,\mathrm{d}\tilde{x}
	+ \frac{1-v^2}{4}\int^{\tilde{y}^{+}}_{\tilde{y}^{-}}\underbrace{\bigg[\hat{h}_{\tilde{y}}^{*}\hat{h}_{\tilde{x}}+\hat{h}_{\tilde{y}}\hat{h}_{\tilde{x}}^{*}\bigg]^{\tilde{x}^{+}}_{\tilde{x}^{-}}}_{I_{1}}\,\mathrm{d}\tilde{y} 
	\\
	&-\frac{iv}{2}\Bigg(
		\int^{\tilde{y}^{+}}_{\tilde{y}^{-}}\underbrace{\bigg[\hat{h}\hat{h}^{*}_{\tilde{y}}\bigg]^{\tilde{x}^{+}}_{\tilde{x}^{-}}}_{I_{2}} \,\mathrm{d}\tilde{y} + \int^{\tilde{x}^{+}}_{\tilde{x}^{-}}\bigg[\hat{h}\hat{h}^{*}_{\tilde{x}}\bigg]^{\tilde{y}^{+}}_{\tilde{y}^{-}} \,\mathrm{d}\tilde{x}
	\Bigg).
\end{aligned} 
\end{equation}
The Galilean transformation is defined under the assumption that the body is moving in the $\tilde{x}$-direction. Under this condition, we demonstrate that $\bar{F}_{T}$ in the $\tilde{y}$-direction is zero due to symmetry in the $\tilde{y}^{\pm}$-direction similar to the observations of \citet{Roh2019bee}. 
Given symmetry about the $\tilde{x}$-axis $(\hat{h}(\tilde{x},\tilde{y}^{+})=\hat{h}(\tilde{x},\tilde{y}^{-}),\hat{h}_{\tilde{x}}(\tilde{x},\tilde{y}^{+})=\hat{h}_{\tilde{x}}(\tilde{x},\tilde{y}^{-}),\hat{h}_{\tilde{y}}(\tilde{x},\tilde{y}^{+})=-\hat{h}_{\tilde{y}}(\tilde{x},\tilde{y}^{-}))$, the arguments of the integrals with respect to $\tilde{x}$ are simply zero. For the $\tilde{y}$ integrals, we can split them up into two integrals,
\begin{equation}
	\frac{1-v^2}{4}\left(\int^{\tilde{y}^{+}}_{0}I_{1}\,\mathrm{d}\tilde{y}
	+\int^{0}_{\tilde{y}^{-}}I_{1}\,\mathrm{d}\tilde{y}\right) 
	-\frac{iv}{2}\left(
	\int^{0}_{\tilde{y}^{-}}I_{2}\,\mathrm{d}\tilde{y}
	+
	\int^{\tilde{y}^{+}}_{0}I_{2} \,\mathrm{d}\tilde{y}\right),
\end{equation}
Where $I_{1},I_{2}$ are denoted by underbraces in \eqref{ap_eq:ythrust_tidy}. Due to symmetry, $\hat{h}_{\tilde{y}}(\tilde{x},\tilde{y})=-\hat{h}_{\tilde{y}}(\tilde{x},-\tilde{y})$, resulting in 
\begin{equation}
	\frac{1-v^2}{4}\left(\int^{\tilde{y}^{+}}_{0}I_{1}\,\mathrm{d}\tilde{y}
	-\int^{\tilde{y}^{+}}_{0}I_{1}\,\mathrm{d}\tilde{y}\right) 
	-\frac{iv}{2}\left(
	\int^{\tilde{y}^{+}}_{0}I_{2}\,\mathrm{d}\tilde{y}
	-
	\int^{\tilde{y}^{+}}_{0}I_{2} \,\mathrm{d}\tilde{y}\right)=0.
\end{equation}
Hence, the thrust in the $\tilde{y}$-direction is
\begin{equation}
	\bar{F}_{T}^{\tilde{y}} = \langle\hat{Q},\hat{h}_{\tilde{y}}\rangle = 0. 
\end{equation}

\section{Power derivation}\label{ap_sec:power}

We begin with the total energy for the two-dimensional wave equation
\begin{equation}
	E(t) = \frac{1}{2}\int^{x^+}_{x^-}\int^{y^+}_{y^-} h_x^2 + h_y^2 + h_{t}^2 \,\mathrm{d}x\,\mathrm{d}y,
\end{equation}
and transform into the Galilean moving frame 
\begin{equation}
	E_{v}(\tilde{t}) = \frac{1}{2}\int^{\tilde{x}^{+}}_{\tilde{x}^{-}}\int^{\tilde{y}^{+}}_{\tilde{y}^{-}}
	\bigg[ (1+v^2)h_{\tilde{x}}^{2} - 2vh_{\tilde{x}}h_{\tilde{t}} + h_{\tilde{t}}^2 + h_{\tilde{y}}^2\bigg]\,\mathrm{d}\tilde{x}\,\mathrm{d}\tilde{y}.
\end{equation}
Taking a derivative with respect to time in the moving frame, we get
\begin{equation}
	\frac{\mathrm{d}E_{v}}{\mathrm{d}\tilde{t}} =  \int^{\tilde{x}^{+}}_{\tilde{x}^{-}}\int^{\tilde{y}^{+}}_{\tilde{y}^{-}}
	(1+v^2)h_{\tilde{x}}h_{\tilde{x}\tilde{t}} + \left(h_{\tilde{t}}-vh_{\tilde{x}} \right)h_{\tilde{t}\tilde{t}}
	-vh_{\tilde{x}\tilde{t}}h_{\tilde{t}} + h_{\tilde{y}}h_{\tilde{y}\tilde{t}} \,\mathrm{d}\tilde{x}\,\mathrm{d}\tilde{y}.
\end{equation}
The $h_{\tilde{t}\tilde{t}}$ term can be substituted for the wave equation in the moving frame \eqref{ap_eq:wave_eq_moving}, such that
\begin{equation}\label{ap_eq:power_before_time_avg}
\begin{aligned}
	\frac{\mathrm{d}E_{v}}{\mathrm{d}\tilde{t}} = \int^{\tilde{x}^{+}}_{\tilde{x}^{-}}\int^{\tilde{y}^{+}}_{\tilde{y}^{-}}&\bigg[(1+v^2)h_{\tilde{x}}h_{\tilde{x}\tilde{t}}  - vh_{\tilde{x}\tilde{t}}h_{\tilde{t}} + h_{\tilde{y}}h_{\tilde{y}\tilde{t}} 
	\\
	&+ (h_{\tilde{t}}-vh_{\tilde{x}})\Big((1-v^2)h_{\tilde{x}\tilde{x}} + 2vh_{\tilde{x}\tilde{t}} + h_{\tilde{y}\tilde{y}} + Q \Big)\bigg]
	\,\mathrm{d}\tilde{x}\,\mathrm{d}\tilde{y}.
\end{aligned}
\end{equation}
The source moves periodically like \eqref{eq:periodic_assumption} and once we time-average, the left-hand side of \eqref{ap_eq:power_before_time_avg} is zero. Rearranging with integration by parts, we arrive at:
\begin{equation}
\begin{aligned}
	\langle\hat{Q},-i\hat{h}\rangle -v\langle\hat{Q},\hat{h}_{\tilde{x}}\rangle = 
	-\frac{1-v^2}{4}\int^{\tilde{y}^{+}}_{\tilde{y}^{-}}\Big[i\hat{h}_{\tilde{x}}\hat{h}^{*} - i\hat{h}_{\tilde{x}}^{*}\hat{h} \Big]^{\tilde{x}^{+}}_{\tilde{x}^{-}}\,\mathrm{d}\tilde{y} 
	-\frac{1}{4}\int^{\tilde{x}^{+}}_{\tilde{x}^{-}}\Big[i\hat{h}_{\tilde{y}}\hat{h}^{*} - i\hat{h}_{\tilde{y}}^{*}\hat{h} \Big]^{\tilde{y}^{+}}_{\tilde{y}^{-}}
	\\
	+v\left(\frac{1}{4}\int^{\tilde{y}^{+}}_{\tilde{y}^{-}}\Big[(1-v^2) |\hat{h}_{\tilde{x}}|^2
	+ |\hat{h}|^2
	- |\hat{h}_{\tilde{y}}|^{2}\Big]^{\tilde{x}^{+}}_{\tilde{x}^{-}}\,\mathrm{d}\tilde{y} +\frac{1}{4}\int^{\tilde{x}^{+}}_{\tilde{x}^{-}}\Big[\hat{h}_{\tilde{x}}^{*}\hat{h}_{\tilde{y}} + \hat{h}_{\tilde{x}}\hat{h}_{\tilde{y}}^{*} \Big]^{\tilde{y}^{+}}_{\tilde{y}^{-}}	\,\mathrm{d}\tilde{x}
	\right).
\end{aligned}
\end{equation}
This is interpreted as an equality between power injected by the source $\hat{Q}$ and the power radiated by the waves. Hence, we define power injected as:
\begin{equation}
	 \overline{\mathrm{Pow}} = \langle\hat{Q},-i\hat{h}\rangle - v\langle\hat{Q},\hat{h}_{\tilde{x}} \rangle.
\end{equation}

The power can be understood further with the prescribed plane wave used in Section \ref{ap_sec:xthrust}:
\begin{equation}\label{ap_eq:power_angles}
\begin{aligned}
	\langle\hat{Q},-i\hat{h}\rangle -v\langle\hat{Q},\hat{h}_{\tilde{x}}\rangle=& \frac{1}{4}\int^{\tilde{y}^{+}}_{\tilde{y}^{-}} \left[\frac{2\cos\alpha - 2v\cos^2\alpha +v^3\cos^2\alpha-v^3-\sin^2\alpha}{(1-v\cos\alpha)^2} |\hat{h}|^2\right]^{^{\tilde{x}^{+}}}_{\tilde{x}^{-}}\,\mathrm{d}\tilde{y}
	\\
	&+\frac{1}{4}\int^{\tilde{x}^{+}}_{\tilde{x}^{-}}\left[\frac{\sin\alpha}{(1-v\cos\alpha)^2}|\hat{h}|^2\right]_{\tilde{y}^{-}}^{\tilde{y}^{+}}\,\mathrm{d}\tilde{x}.
\end{aligned}
\end{equation}
If we have a wave coming out of the raft at the front and back ($\alpha=0,\pi$ for $\tilde{x}^{+},\tilde{x}^{-}$) in the subcritical case, the power expression will collapse down to the same sum of fore aft waves like in one dimension \citep{odonovan2026},
\begin{equation}
	\overline{\mathrm{Pow}} = 
	\frac{1}{2}\int^{\tilde{y}^{+}}_{\tilde{y}^{-}}\bigg[ \frac{1}{1-v}|\hat{h}_{+}|^2 + \frac{1}{1+v} |\hat{h}_{-}|^2 \bigg]\,\mathrm{d}\tilde{y}. 
\end{equation}
We also note that since $\sin\alpha$ and $\cos\alpha$ are antisymmetric and symmetric about the $\tilde{x}$-axis, respectively, the integrand in the last term of \eqref{ap_eq:power_angles} amounts to a sum of equal left and right waves. This is expected due to the symmetry in the $\tilde{y}$-direction.
And similar results can be found to relate back to the one-dimensional supercritical case using \eqref{ap_eq:sup_angle_1} and \eqref{ap_eq:sup_angle_2}.

\section{Variational calculus optimisation on start-up}\label{ap_sec:start-up_norm}

The start-up wave field \eqref{eq:startup_greens} is substituted into the action
\begin{equation}
	f = \bar{F}_T + \lambda C,
\end{equation}
where $\lambda$ is a Lagrange multiplier and $C$ is the bounded norm constraint, giving:
\begin{equation}
	\label{ap_eq:norm_action_full}
	\begin{gathered}
		f = \frac{i}{16}\Bigg[-\int^{\frac{l}{2}}_{-\frac{l}{2}}\int^{\frac{l}{2}}_{-\frac{l}{2}}\hat{Q}^{*}(x,y)\left(\int^{\frac{l}{2}}_{-\frac{l}{2}}\int^{\frac{l}{2}}_{-\frac{l}{2}}\hat{Q}(X,Y)\frac{x-X}{|\boldsymbol{x}-\boldsymbol{X}|}H^{(1)}_{1}(|\boldsymbol{x}-\boldsymbol{X}|)\,\mathrm{d}X\,\mathrm{d}Y\right)\,\mathrm{d}x\,\mathrm{d}y 
		\\
		+\int^{\frac{l}{2}}_{-\frac{l}{2}}\int^{\frac{l}{2}}_{-\frac{l}{2}}\hat{Q}(x,y)\left(\int^{\frac{l}{2}}_{-\frac{l}{2}}\int^{\frac{l}{2}}_{-\frac{l}{2}}\hat{Q}^{*}(X,Y)\frac{x-X}{|\boldsymbol{x}-\boldsymbol{X}|}H^{(2)}_{1}(|\boldsymbol{x}-\boldsymbol{X}|)\,\mathrm{d}X\,\mathrm{d}Y\right)\,\mathrm{d}x\,\mathrm{d}y\Bigg]
		\\
		+ \lambda\int^{y^{+}}_{y^{-}}\int^{x^{+}}_{x^{-}} \hat{Q}^{*}\hat{Q}\,\mathrm{d}x\,\mathrm{d}y.
	\end{gathered}
\end{equation}
We apply a perturbation of the form,
\begin{equation}\label{ap_eq:perturb_defn}
    \delta f = \lim_{\epsilon \rightarrow0}\frac{f\left(\hat{Q}({x}) + \epsilon\zeta({x})\right) - f\left(\hat{Q}({x})\right)}{\epsilon}.
\end{equation}
Swapping the order of integration in the resulting perturbed action, we 
follow the same process as \citep{odonovan2026} by setting the perturbation to be zero. This results in the following condition:
\begin{equation}\label{ap_eq:norm_results_before_L}
-\frac{i}{8} \int^{\frac{l}{2}}_{-\frac{l}{2}}\int^{\frac{l}{2}}_{-\frac{l}{2}}\hat{Q}(X,Y)\frac{x-X}{|\boldsymbol{x}-\boldsymbol{X}|}J_{1}(|\boldsymbol{x}-\boldsymbol{X}|)\,\mathrm{d}X\,\mathrm{d}Y + \lambda \hat{Q}(x,y) = 0. 
\end{equation}
The first term is in the null space of the operator $\mathcal{L}_{2D} = \partial_{xx} +\partial_{yy}+\mathbb{I}$. Therefore, applying the operator $\mathcal{L}_{2D}$ results in the optimal condition:
\begin{equation}\label{eq:norm_cond}
	\mathcal{L}_{2D}\hat{Q} = 0.
\end{equation}
We will solve this equation with separation of variables of the form $\hat{Q}(x,y) = X(x)Y(y)$, splitting \eqref{eq:norm_cond} into two ODEs
\begin{align}
	\frac{X''(x)}{X(x)} = -\kappa^2,
	&&
	\frac{Y''(y)}{Y(y)} = \kappa^2 - 1,
\end{align}
where $\kappa\in\mathbb{R}$ is a separation constant. The solution is 
\begin{equation}\label{ap_eq:norm_solved_Q}
	\hat{Q}(x,y) =  A e^{i\kappa x+i\sqrt{1-\kappa^2}y} + B e^{-i\kappa x+i\sqrt{1-\kappa^2}y} + C e^{i\kappa x-i\sqrt{1-\kappa^2}y} + D e^{-i\kappa x-i\sqrt{1-\kappa^2}y}, 
\end{equation}
where $A,B,C,D\in\mathbb{C}$ are constants.
It is not simple to find our constants as it was in the one-dimensional case (e.g. by inserting \eqref{ap_eq:norm_solved_Q} into \eqref{ap_eq:norm_results_before_L} and the norm constraint \eqref{eq:norm}). Instead, we can apply \eqref{ap_eq:norm_solved_Q} as a constraint to the optimisation to find the constants and validate the methods. This is referred to as the analytically informed optimisation.

\section{Subcritical boundary condition}\label{ap_sec:sub_bc}

Finding general subcritical boundary conditions is a non-trivial task.
The following approach exploits the invariance the wave equation under a Lorentz transformation, allowing us to draw inspiration from the start-up case and use the same Green's function \eqref{eq:startup_greens}.  Once the solution is found, we can transform back to the rest frame, giving the solution for a source moving through the rest frame. A Galilean transformation is then taken to work in the moving frame of interest to infer the correct boundary conditions.

\subsection{Finding our wave field}

We will begin with the wave equation in the dimensional set-up with a time-periodic translating point source, $Q = \delta(\tilde{x})\delta(\tilde{y})\mathrm{e}^{-i\omega \tilde{t}}$ and we focus on the real part of the source and its resulting wave field throughout. 
Applying the Lorentz transformation \eqref{eq:lorentz_transform} results in:
\begin{equation}\label{eq:PDE_lorentz}
	\frac{1}{c^2}h^{\prime}_{t't'} - \nabla^{'2}h^{\prime} = \delta\left(\frac{x'}{\gamma}\right)\delta(y')e^{-i\omega\gamma(t'+\frac{U}{c}x')},
\end{equation}
where the prime denotes Lorentz coordinates as opposed to differentiation.
Separating the spatial and temporal parts, $h^{\prime}(x^{\prime},y^{\prime},t^{\prime}) = \hat{h}^{\prime}(x^{\prime},y^{\prime})\mathrm{e}^{-i\omega\gamma t^{\prime}}$, 
and our PDE becomes
\begin{equation}
	{\lambda}^2\hat{h}^{\prime} + \nabla^{{\prime}2}\hat{h}^{\prime} = -\delta\left(\frac{x^{\prime}}{\gamma}\right)\delta\left(y^{\prime}\right)\mathrm{e}^{-i\omega\gamma\frac{U}{c^2}x^{\prime}},
\end{equation}
where $\lambda=\omega\gamma/c$. Using the Green's function for the Helmholtz equation, the wave field is:
\begin{equation}
	\hat{h}^{\prime} = \frac{i}{4}\int^{\infty}_{-\infty}\int^{\infty}_{-\infty} H^{(1)}_{0}\Big(\lambda |\boldsymbol{x^{\prime}}- \boldsymbol{X^{\prime}}|\Big)\delta\left(\frac{X^{\prime}}{\gamma}\right)\delta\left(Y\right)\mathrm{e}^{-i\omega\gamma\frac{U}{c^2}X^{\prime}}\,\mathrm{d}X^{\prime}\,\mathrm{d}Y^{\prime},
\end{equation}
and a change of variables, $\xi=X^{\prime}/\gamma$, is used to simplify the integrals, such that
\begin{equation}
	\hat{h}^{\prime} = \frac{i\gamma}{4}H^{(1)}_{0}\left(\lambda|\boldsymbol{x'}|\right)\mathrm{e}^{0}.
\end{equation}
This is the wave field in the Lorentz moving frame. Taking the inverse Lorentz transformation, the time dependent $h$ in the untransformed frame is:
\begin{equation}\label{eq:sol_back_at_rest}
	h^{\mathrm{rest}} = \frac{i\gamma}{4}H_{0}^{(1)}\left(\frac{\omega\gamma}{c}\left(\gamma^2(x-Ut)^2+y^2\right)^{\frac{1}{2}}\right)\mathrm{e}^{-i\omega\gamma^2\left(t-\frac{U}{c^2}x\right) }.
\end{equation}
The frame we are interested in is the Galilean moving frame. This requires another coordinate system which will be denoted with a tilde $\tilde{\boldsymbol{x}}$. The dimensional transformation,
\begin{align}
	\tilde{x} = x-Ut, && \tilde{y} = y, && \tilde{t} = t,
\end{align}
is applied to \eqref{eq:sol_back_at_rest} producing the following wave field in the Galilean moving frame:
\begin{equation}
	h = \frac{i\gamma}{4}H^{(1)}_{0}\left(\frac{\omega\gamma}{c}(\gamma^2\tilde{x}^2+\tilde{y}^2)^{\frac{1}{2}}\right)\mathrm{e}^{i\omega\gamma^2\frac{U}{c^2}\tilde{x}}\mathrm{e}^{-i\omega \tilde{t}}.
\end{equation}
Removing the periodic time component results in the dimensional wave field $\hat{h}(\tilde{x},\tilde{y})$ in the Galilean frame,
\begin{equation}\label{eq:gal_transform_sol}
	\hat{h}(\tilde{x},\tilde{y}) = \frac{i\gamma}{4}H^{(1)}_{0}\left(\frac{\omega\gamma}{c}(\gamma^2\tilde{x}^2+\tilde{y}^2)^{\frac{1}{2}}\right)\mathrm{e}^{i\omega\gamma^2\frac{U}{c^2}\tilde{x}}.
\end{equation}

We use $\hat{h}$ to find a far-field boundary condition. In the start-up case, we began with,
\begin{equation}\label{eq:base_bc_form}
	\boldsymbol{\hat{n}}\cdot\nabla h = \boldsymbol{\hat{n}}\cdot i\boldsymbol{k}h,
\end{equation}
for unit normal vector $\boldsymbol{\hat{n}}$. 
We will find a new boundary condition by evaluating the left--hand side explicitly, and begin by calculating $\tilde{\nabla}\hat{h}$:
\begin{equation}\label{ap_eq:grad_h_tilde}
	\tilde{\nabla}\hat{h} = 
	\begin{pmatrix}
		\hat{h}_{\tilde{x}} \\
		\hat{h}_{\tilde{y}}
	\end{pmatrix}	
	= \begin{pmatrix}
		\frac{i\omega\gamma^2U}{c^2}\hat{h} 
		- \frac{\omega\gamma}{c}\frac{\gamma^2\tilde{x}}{\sqrt{\gamma^2\tilde{x}^2+\tilde{y}^2}}
		\frac{i\gamma}{4}
		H^{(1)}_{1}\left(\frac{\omega\gamma}{c}(\gamma^2\tilde{x}^2+\tilde{y}^2)^{\frac{1}{2}}\right)
		\mathrm{e}^{i\omega\gamma^2\frac{U}{c^2}\tilde{x}}
		\\
		\\
		-\frac{\omega\gamma}{c}\frac{\tilde{y}}{\sqrt{\gamma^2\tilde{x}^2 + \tilde{y}^2}}\frac{i\gamma}{4}H^{(1)}_{1}\left(\frac{\omega\gamma}{c}(\gamma^2\tilde{x}^2+\tilde{y}^2)^{\frac{1}{2}}\right)\mathrm{e}^{i\omega\gamma^2\frac{U}{c^2}\tilde{x}}
	\end{pmatrix}.
\end{equation}
For large $z$, $H_{0}^{(1)}(z) = -iH_{1}^{(1)}(z)$, correct to first order. This simplifies \eqref{ap_eq:grad_h_tilde} to an approximate form in the far-field:
\begin{equation}
	\tilde{\nabla}\hat{h} \approx
	\begin{pmatrix}
		\frac{i\omega\gamma^2U}{c^2}\hat{h} 
		+ \frac{i\omega\gamma^2}{c}\frac{\gamma\tilde{x}}{\sqrt{\tilde{x}^2\gamma^{2} + \tilde{y}^2}}\hat{h}
		\\
		\\
		\frac{i\omega\gamma}{c}\frac{\tilde{y}}{\sqrt{\tilde{x}^2\gamma^2 + \tilde{y}^{2}}}\hat{h}
	\end{pmatrix}.
\end{equation}
The normal vector will be written similarly to the start-up case where
\begin{equation}
	\boldsymbol{\hat{n}} = 
	\begin{pmatrix}
		\cos(\tilde{\theta}) 
		\\
		\sin(\tilde{\theta})
	\end{pmatrix},
\end{equation}
and $\tilde{\theta}$ is the effective angle. 
To simplify the boundary condition, $\hat{h}_{\tilde{x}}$ is scaled by $\gamma^{-1}$ and the result is as follows:
\begin{equation}
	\underline{\tilde{n}} \cdot 
	\begin{pmatrix}
		\gamma^{-1} \hat{h}_{\tilde{x}}
		\\
		\hat{h}_{\tilde{y}}
	\end{pmatrix} 
	= \cos(\tilde{\theta})\left(\frac{i\omega\gamma U}{c^2}\hat{h} 
		+ \frac{i\omega\gamma}{c}\frac{\gamma\tilde{x}}{\sqrt{\tilde{x}^2\gamma^{2} + \tilde{y}^2}}\hat{h}\right) 
		+ \sin(\tilde{\theta})\left(	\frac{i\omega\gamma}{c}\frac{\tilde{y}}{\sqrt{\tilde{x}^2\gamma^2 + \tilde{y}^{2}}}\hat{h}\right).
\end{equation}
The relation between our coordinates and the angle $\tilde{\theta}$ presents itself naturally as:
\begin{align}\label{eq:eff_angle_relations}
	\cos(\tilde{\theta}) = \frac{\gamma\tilde{x}}{\sqrt{\gamma^2\tilde{x}^2 + \tilde{y}^2}}, &&
	\sin(\tilde{\theta}) = \frac{\tilde{y}}{\sqrt{\gamma^2\tilde{x}^2 + \tilde{y}^2}}.
\end{align}
The dimensional subcritical boundary condition is found to be:
\begin{equation}\label{ap_eq:simplified_v_bc}
	\frac{1}{\gamma}\cos(\tilde{\theta})\hat{h}_{\tilde{x}} + \sin(\tilde{\theta})\hat{h}_{\tilde{y}} = 
	\frac{i\omega\gamma}{c}\hat{h}\left(1 + v\cos(\tilde{\theta})\right).
\end{equation}
The solution \eqref{eq:gal_transform_sol} is the same as that of \citet{barucq2022_boundary_conds} found using a Prandtl-Glauert-Lorentz transformation, albeit with a different sign convention evident in the exponential. While the outgoing plane wave boundary conditions appear to differ slightly and are evaluated on an elliptical boundary, the fact that they satisfy the same solution suggests they are equivalent but in different set-ups. 

\newpage
\bibliographystyle{jfm}
\bibliography{bibfile}

\end{document}